\documentclass[]{aa}  
\usepackage{graphicx}
\usepackage{txfonts}
\usepackage{amsmath}
\usepackage{color}
\usepackage[colorlinks,breaklinks]{hyperref}
\hypersetup{linkcolor=blue,citecolor=blue,filecolor=black,urlcolor=blue}
\usepackage{amssymb}
\usepackage{comment}
\usepackage{hhline}
\usepackage[normalem]{ulem}
\usepackage{csvsimple}
\usepackage{tabularx}
\usepackage{mathtools}
\usepackage{booktabs}

\usepackage[]{lineno} 

\newcommand{\sep}{\pkg{SEP}}
\newcommand{\ici}{\ensuremath{i.c.i.}{}}

\newcommand{\pkg}[1]{\textsc{\texttt{#1}}}

\newcommand{\LCDM}{$\Lambda$CDM }
\newcommand{\prob}[2]{\mathcal{P}\left( #1 \mid #2\right)}

\begin{document}

\title{Accuracy of environmental tracers and consequence for determining the Type Ia Supernovae magnitude step}

\titlerunning{Accuracy of environmental tracers and consequence for determining the Type Ia Supernovae magnitude step}
\authorrunning{M.~Briday et al.}

\author{M.~Briday \inst{\ref{ipnl}},
M.~Rigault \inst{\ref{ipnl}},
R.~Graziani \inst{\ref{ipnl}}, 
Y.~Copin \inst{\ref{ipnl}}, 
G.~Aldering \inst{\ref{Berkeley1}},
M.~Amenouche \inst{\ref{lpc}},
V.~Brinnel \inst{\ref{Berlin}},
A.~G.~Kim \inst{\ref{Berkeley1}}, 
Y.-L.~Kim \inst{\ref{ipnl}}, 
J.~Lezmy \inst{\ref{ipnl}},
N.~Nicolas \inst{\ref{ipnl}},
J.~Nordin \inst{\ref{Berlin}},
S.~Perlmutter \inst{\ref{Berkeley2}}, 
P.~Rosnet \inst{\ref{lpc}}, 
M.~Smith \inst{\ref{ipnl}}
}


\institute{
    Univ Lyon, Univ Claude Bernard Lyon 1, CNRS, IP2I Lyon / IN2P3, IMR 5822, F-69622, Villeurbanne, France
    \label{ipnl}
\and
    Universit\'e Clermont Auvergne, CNRS/IN2P3, Laboratoire de Physique
    de Clermont, F-63000 Clermont-Ferrand, France \label{lpc}
\and
    Lawrence Berkeley National Laboratory, 1 Cyclotron Rd., Berkeley, CA, 94720, USA \label{Berkeley1}
\and
    Department of Physics, University of California Berkeley, 366 LeConte Hall MC 7300, Berkeley, CA 94720-7300, USA \label{Berkeley2}
\and
    Institute of Physics, Humboldt-Universität zu Berlin, Newtonstr. 15, 12489 Berlin, Germany \label{Berlin}
}

\date{}

\abstract{
Type Ia Supernovae (SNe~Ia) are standardizable candles that allow us to measure the recent expansion rate of the Universe. Due to uncertainties in progenitor physics, potential astrophysical dependencies may bias cosmological measurements if not properly accounted for. 
The dependency of the intrinsic luminosity of SNe~Ia with their host-galaxy environment is often used to standardize SNe~Ia luminosity and is commonly parameterized as a step function. This functional form implicitly assumes two-populations of SNe~Ia. In the literature, multiple environmental indicators have been considered, finding different, sometimes incompatible, step function amplitudes.
We compare these indicators in the context of a two-populations model, based on their ability to distinguish the two populations. We show that local H$\alpha$-based specific star formation rate (lsSFR) and global stellar mass are better tracers than, for instance, host galaxy morphology. 
We show that tracer accuracy can explain the discrepancy between the observed SNe~Ia step amplitudes found in the literature. Using lsSFR or global mass to distinguish the two populations can explain all other observations, though lsSFR is favoured.
As lsSFR is strongly connected to age, our results favour a prompt and delayed population model. In any case, there exists two populations that differ in standardized magnitude by at least $0.121\pm0.010\,\mathrm{mag}$.
}


\maketitle

\section{Introduction}
\label{sec:introduction}

Type Ia supernovae (SNe~Ia) are powerful empirically standardized distance indicators. They enabled the discovery of the acceleration of the expansion of the Universe \citep{riess1998, perlmutter1999}, and remain today key cosmological probes in the context of the new generation of surveys \citep{scolnic2019}.
SNe~Ia play an important role to probe the nearby Universe ($z<0.3$) and are the last step of the direct distance ladder to derive the Hubble–Lemaître constant $H_0$ \citep[e.g.][]{freedman2001,riess2009}. 
Interestingly, when calibrating the SNe~Ia absolute luminosity using the Cepheid period-luminosity relation, this direct $H_0$ measurement is $4.4\sigma$ higher than expectation based on the \LCDM model anchored by \cite{planck2018} data \citep{riess2019,reid2019}. This ``tension'' has received a lot of attention as it could be a sign of new fundamental physics \citep{knox2020}. This finding is supported by analyses of strongly lensed quasars that are also reporting high $H_0$ measurements \citep[e.g.,][]{wong2020}. However \cite{freedman2019} find a lower $H_0$ value when using tip of the red giant branch (TRGB) distances in place of the Cepheids.

This raises the question of systematic uncertainties affecting direct $H_0$ measurements in particular, and the distances derived from the observation of the SNe~Ia in general. \cite{rigault2015} suggests that an unaccounted for astrophysical bias, affecting the derivation of the absolute SNe~Ia luminosity, could explain at least part of the tension. Indeed, SNe~Ia from the calibrating sample significantly differ from the Hubble flow ones as they are selected such that their host galaxy also contains Cepheid and are thus star forming. \cite{rigault2020} claim that SNe~Ia from younger environments are $0.16$~mag fainter than those from older environments, leading to a bias on $H_0$ because of the aforementioned selection effect. Yet, \cite{riess2019} have mimicked the Cepheid selection function onto the Hubble flow sample and find no variation in $H_0$, suggesting that they are not affected by this astrophysical effect (see also \cite{jones2015}).

After more than a decade of analyses, the amplitude and the root causes of the astrophysical biases affecting the distance measurements from SNe~Ia remains unclear.

Early Ia rates studies have shown evidences that two populations of SNe~Ia may exist: one arising from young (<100~Myr) progenitor systems and one related to older (Gyr), most evolved progenitors \citep[i.e. the A+B or prompt/delayed models, see e.g.,][]{mannucci2005,mannucci2006, scannapieco2005, sullivan2006, aubourg2008, smith2012, maozmannucci2014, rodney2014}. But the first significant evidence of an astrophysical bias in the SN distance derivations was the observed dependency of the standardized SNe~Ia magnitude (using classical 2-parameters light curve standardization method) with host galaxy stellar mass \citep[e.g.][]{kelly2010,sullivan2010,lampeitl2010,gupta2011,childress2013b,betoule2014,uddin2017,ponder2020}: SNe~Ia from massive galaxies ($\mathrm{M_*>10^{10}M_\odot}$) are brighter, after standardisation, by $\sim0.1$~mag. 

We use the term ``magnitude-step'' to describe the difference in average standardized magnitudes between two SN~Ia sub-samples defined from an environmental tracer cut-off. This simple functional form has been shown by \cite{childress2013b} to be the best fit to data in comparison to a linear trend or other theoretically inspired forms. We further highlight that, in practice, this data-driven “step” implies that there exists two populations of “SN~Ia+environment” that are simultaneously present; this observation is a central point of this paper.

The term ``mass-step'' has been extensively used in the literature for the global host-stellar mass tracer. The mass-step is used as a third standardisation parameter in many recent SNe~Ia cosmological analyses \citep{sullivan2010,betoule2014,scolnic2018}, including the direct $H_0$ measurements from \cite{riess2016,riess2019}. The amplitude of this effect is $\sim 0.08$~mag. Yet, the underlying physics causing this magnitude dependency remains unclear, and so is the proper way to account for such astrophysical biases. 

In the last decade, many host environmental studies seem to converge towards either the age of the progenitor or dust around the progenitor or in the host interstellar medium as the origin of the mass-step. \citet{rigault2013,rigault2020,roman2018,kim2018,kelsey2021} would suggest age, while others, like \cite{brout2021}, suggest that variable dust extinction curves affecting the observed color of the supernova can explain correlations with host galaxy properties. \cite{rigault2020} show the most significant correlation between the luminosity of the SN and the properties of the environment. They split their SNe~Ia as a function of the specific Star Formation Rate (sSFR) derived from H$\alpha$ flux measured within a 1~kpc radius projected onto the local environment around the SN (local sSFR; lsSFR). The SNe~Ia having a large lsSFR, hence a large fraction of young stars in their vicinity, are fainter than those from passive local environments by $0.163\pm0.029$~mag.
Since high-mass galaxies favor older stellar populations, massive hosts favor SNe from old environments, and so are, on average, brighter after light-curve standardization; resulting in the mass-step.

Surprisingly, while most SN samples now observe significant correlation between host properties and standardized SN magnitudes, the magnitude step amplitudes differ and seem incompatible. For instance, the SNLS-5 years and SDSS data from \cite{roman2018}, updating the JLA catalog from \cite{betoule2014}, find a local U-V step of $0.091\pm0.013$~mag, seemingly incompatible with \cite{rigault2020}. \cite{jones2018}, using the low-redshift bin of the Pantheon dataset \citep{scolnic2018}, reported that locally massive environments, i.e. having a large surface density of stars, are fainter by $0.067\pm 0.017$~mag. While in agreement with the local mass-step reported in \citet[][$0.059\pm0.024$~mag]{rigault2020}, this effect is significantly weaker than that observed using the lsSFR indicator. Finally, using the Pantheon dataset \cite{pruzhinskaya2020} found that SNe~Ia from elliptical and lenticular galaxies are brighter ($0.058\pm0.019$~mag) in agreement with \citet[][using JLA, $0.018\pm0.052$~mag]{kim2019}, and \citet[][$0.04\pm0.05$~mag]{henne2017}.

As already pointed out by \cite{jones2018}, this variety of results, made using different host tracers, local or global, brings confusion about how to best account for astrophysical biases in SN cosmology, and notably on the derivation of $H_0$. In this paper, we try to clarify this situation by studying how well each environmental indicator is able (or not) to trace a given environmental property. 

We start, in Section~\ref{sec:concept}, by presenting that, mathematically, if two SN~Ia populations were to exist, the observed amplitude of their true standardized magnitude difference linearly depends on the ability of a tracer to accurately measure which population a SN belongs to. We present in Section~\ref{sec:data} the data sample we use for this work and we describe the methodology used to extract the aforementioned environmental tracers' measurements. Then, we apply in Section~\ref{sec:cont_from_lssfr} our ``two-populations'' model on these data and we present our findings in Section~\ref{sec:results}, comparing with results from the literature. For this we used the H$\alpha$-based lsSFR tracer as reference, and we test this hypothesis in Section~\ref{sec:discussion}. We discuss our findings and we conclude in Section~\ref{sec:conclusion}.

\section{The ``two-populations'' model}
\label{sec:concept}

As discussed in the introduction, many SN cosmological analyses use ``step'' functions to account for environmental dependencies in the derivation of distances. These steps are the difference of average properties, say ``$q$'', between two sides of a boundary --~``$t_\mathrm{cut}$''~-- in a considered environmental property ``$t$''. For example, the previously mentioned ``mass-step'' is the difference of absolute magnitude ($q$ is the magnitude) of SNe~Ia from low- and high-mass hosts ($t$ is the host stellar mass), conventionally split at the host stellar mass of $t_\mathrm{cut}=10^{10}\;\mathrm{M}_\odot$.

The underlying assumptions, when using a step function, are the following: (1) there exists two categories of SNe, say ``$a$'' and ``$b$'', that differ on average in $q$ and (2) the tracer $t$ is able to probe these categories. Consequently, the amplitude of the observed ``step'' depends on the intrinsic SN properties $q$ and the quality of the tracer $t$.

In the following subsections, we describe our statistical model starting from an illustrative mock example, firstly explaining the mathematical concept without measurement errors, and then including them in the model, to finally introduce the probability function. A detailed mathematical derivation is given Appendix~\ref{app:mathematical_derivation}.

\subsection{Concept of contamination}
\label{sec:conceptnoerr}

Let us assume that two SN populations $a$ and $b$ have a normally distributed quantity $q$, say $\mathcal{N}_a=\mathcal{N}(\mu_a, \sigma_a)$ and $\mathcal{N}_b=\mathcal{N}(\mu_b, \sigma_b)$, for which they differ on average by $\gamma_0=\mu_a - \mu_b$; this difference corresponds to the ``true step amplitude''.

We now assume we have access to a tracer $t$ that is able to discriminate between the $a$ and $b$ populations, but not with perfect accuracy; that is, using the statistical binary classification terminology, with neither perfect specificity nor sensitivity. The tracer classification is based on the cut-off value $t_\mathrm{cut}$ such that the SNe are classified as $a$ or $b$ if they are either above or below the cut, respectively.

Even assuming this tracer provides error-free measurements, we expect misclassification from the tracer inaccuracy: some SNe from the $a$ category will be measured below the $t_\mathrm{cut}$ and will thus be wrongly classified as $b$, and vice versa.

\begin{figure}
    \centering
    \includegraphics[width=\linewidth]{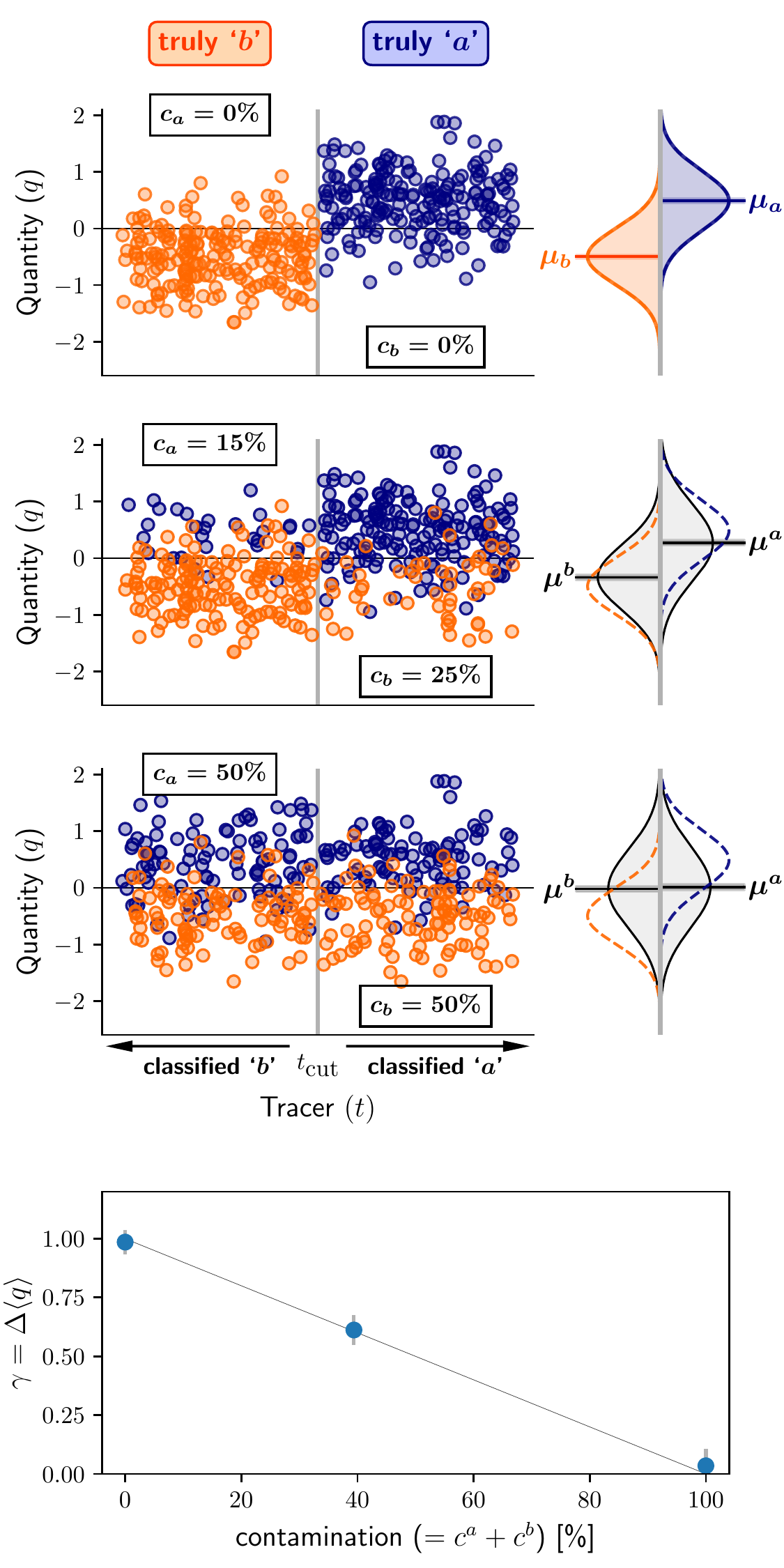}
    \caption{Concept of tracer purity and its impact on the step measurement of an arbitrary quantity $q$. 
    \emph{Top three panels}: 400 mock data, 200 of category $a$ (in blue), normally distributed over the quantity $q$ by $\mathcal{N}(\mu_a=0.5, \sigma_a=0.5)$; and 200 of category $b$ (in orange) with a similar distribution $\mathcal{N}(\mu_b=-0.5, \sigma_b=0.5)$. The figure shows the mock $q$ values as a function of the tracer values $t$ for three different tracers. The data are classified as $a$ or $b$ given the tracer values if they are measured above or below the $t_\textrm{cut}$ cut-off, respectively. Top, middle and bottom panels show tracers with perfect, medium and null ability to track the two categories. Their corresponding right panels show the tracer estimation of the $q$ distribution for each tracers (in grey); to be compared to the true $q$ distribution for each categories (colored dashed lines).
    \emph{Bottom panel}: Evolution of the observed step $\gamma = \mu^a - \mu^b$, the difference of mean $q$ values for targets classified as $a$ or $b$ by the tracers, as a function of the tracer contamination $c = c^a + c^b$;  the markers show each of the three simulations illustrated in the top panels, while the line shows the prediction from Eq.~\ref{eq:stepbias}.}
    \label{fig:contconcept}
\end{figure}

This is illustrated in Fig.~\ref{fig:contconcept} for three cases with varying accuracy: perfect, medium and null. This figure also illustrates how the estimation of the underlying $q$ distributions for each $a$ and $b$ categories is affected by the tracer's inaccuracy, and consequently, how the derived steps are potentially underestimated.
As the tracer accuracy degrades and misclassification cases increase, the number of $a$ category SNe wrongly classified by the tracer as $b$ increases (blue markers in the left-part of the figure); similarly, the fraction of $b$ category SNe misclassified as $a$ increases too (orange markers in the right-part of the panels).
As a consequence of misclassifications, the measured distributions of $q$ for each of the inferred $a$ and $b$ populations broaden and their means converge, so that the step measured using an inaccurate tracer is systematically smaller than the true step.

\subsection{Notations and definitions}
\label{sec:notations}

For clarity, we set here our definitions and nomenclatures in the large-number limit. $N$ is the number of targets, the subscript ${}_x$ denotes true conditions ($x = \{a;b\}$) and the superscript ${}^x$ denotes actual classifications by a tracer. Hence, $N_a^b$ is the number of targets that are truly $a$ but classified as $b$. Accordingly, $N^b = N_a^b + N_b^b$ is the number of targets classified as $b$ by a tracer, and $N_b = N_b^b + N_b^a$ is the number of targets which are intrinsically $b$.

We call contamination the fraction of targets for which the tracer classification differs from the truth. It could either be defined as the fraction $c_a \equiv N_a^b/N_a$ of truly $a$ targets classified as $b$ (resp. $c_b \equiv N_b^a/N_b$); or as the fraction $c^a \equiv N_b^a/N^a$ of classified $a$ targets that actually truly are $b$ (resp. $c^b \equiv N_a^b/N^b$). The two definitions are related as $c^a = N_b/N^a\,c_b$ and $c^b = N_a/N^b\,c_a$. Note also that $1-c_x \equiv N_x^x/N_x$ and $1-c^x \equiv N_x^x/N^x$, where $x$ is either $a$ or $b$. 
\footnote{Following the standard binary classification terminology, if $a$ is the positive condition and the $b$ negative one, then, $N_a$ are the real positive cases ($\textrm{P}$), $N_b$ the real negative ones  ($\textrm{N}$), $N_a^a$ are the true positives ($\textrm{TP}$) and $N_b^b$ are the true negatives ($\textrm{TN}$). Thus $N_b^a$ are the false positives ($\textrm{FP}$) and $N_a^b$ the false negatives ($\textrm{FN}$). Finally $c_a$ is the false negative rate ($\textrm{FNR}$) and $c_b$ is the false positive rate ($\textrm{FPR}$); $c^a$ is the false discovery rate ($\textrm{FDR}$) and $c^b$ is the false omission rate ($\textrm{FOR}$).}

The probability $p^a$ of a target tracer $t_i$ to be measured above the tracer threshold $t_{\mathrm{cut}}$, thus classified as $a$, is the sum of (1) the probability that a target truly is $a$ and properly identified as $a$, and (2) the probability that it truly is $b$ but misclassified as $a$: 
\begin{align}
    \label{eq:p^a}
    p^a &\equiv \frac{N^a}{N} = \frac{N_a^a}{N} + \frac{N_b^a}{N} = (1-c_a)\times\frac{N_a}{N} + c_b\times\frac{N_b}{N} \nonumber \\
    &= (1-c_a) \times p_a + c_b \times (1-p_a)
\end{align}
where $p_b \equiv N_b/N = 1 - N_a/N \equiv 1 - p_a$. Similarly:
\begin{align}
    \label{eq:p^b}
    p^b &= c_a \times p_a + (1-c_b)\times(1-p_a).
\end{align}

While $\mu_a$ and $\mu_b$ are the true means of the SN from categories $a$ and $b$ respectively (see Section~\ref{sec:conceptnoerr}), $\mu^a$ and $\mu^b$ are the distribution means of the considered distributions for each group classified by the tracer (the filled grey distributions in Fig.~\ref{fig:contconcept}). Accordingly, the observed amplitude step $\gamma = \mu^a - \mu^b$, as measured by a tracer, is related to the true intrinsic step $\gamma_0 = \mu_a - \mu_b$ by:
\begin{align}
    \label{eq:stepbias}
    \gamma &= \left( \left(1-c^a\right) \mu_a + c^a \mu_b \right) - 
           \left( c^b \mu_a + \left(1-c^b\right) \mu_b \right) \nonumber\\
           &= \gamma_0 \times \left[ 1- \left( c^a + c^b \right)\right].
\end{align}
This prompts us to define the (total) contamination of a tracer as $c = c^a + c^b$. The linearly decreasing relation in the bottom panel of Fig.~\ref{fig:contconcept} illustrates this equation. 

\subsection{Reference, comparison tracers and measurement errors}
\label{sec:ref_tracer}
\label{sec:concepterr}

It is unlikely that one has access to the true population classification, but rather has to rely on a ``reference tracer'' with respect to which the other tracers will be compared to. This reference tracer is itself an observable associated to its own contamination parameters $c_a^\mathrm{ref}$ and $c_b^\mathrm{ref}$; however, in this analysis, we will generally consider perfectly accurate reference tracer, i.e. $c_a^\mathrm{ref} = c_b^\mathrm{ref}= 0$. The correlation of any other tracer with respect to this \emph{reference} tracer enables to derive the contamination of this \emph{comparison} tracer.

\begin{figure}
    \centering
    \includegraphics[width=\linewidth]{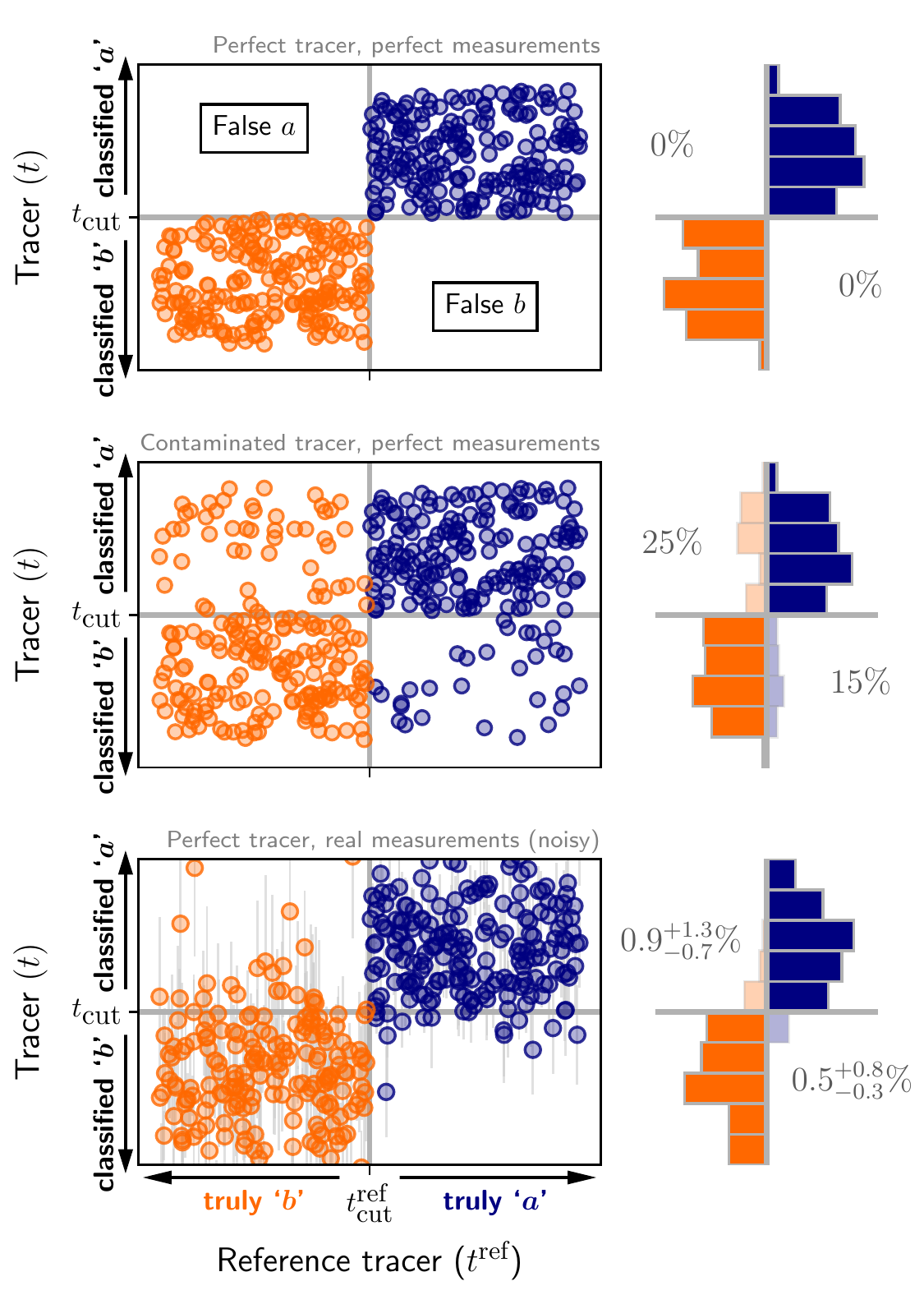}
    \caption{Correlation between uniform mock reference and comparison tracers. The color indicates the true classification following the color-code of Fig.~\ref{fig:contconcept}: blue for $a$, orange for $b$. The right-panel of each row shows the marginalized distribution of each quadrant, following the marker color-code. The light-color histograms correspond to the off-diagonal terms: ``false~$a$'' in light-orange, ``false~$b$'' in light-blue.
    \emph{Top panels:} perfect and noise-free tracer, i.e. without any off-diagonal points; \emph{middle panels:} imperfect ($c_a=15\%$ and $c_b=25\%$) yet noise-free tracer, the sum of its off-diagonal terms corresponds to its contamination parameters; \emph{bottom panels:} perfect but noisy tracer. In this case, plain off-diagonal fractions are $11\%$ in false $a$ and $7.5\%$ in false $b$, while both should be $0\%$ for a perfect tracer; the likelihood estimator (Eq.~\ref{eq:likelihood}, see Section~\ref{sec:measuringcyco}), properly accounting for the fraction of off-diagonal terms caused by measurement errors, provides contamination estimates $c_a = 0.5^{+0.8}_{-0.3}\%$ and $c_b = 0.9^{+1.3}_{-0.7}\%$, compatible with zero.
    }
    \label{fig:corrconcept}
\end{figure}

This is illustrated in Fig.~\ref{fig:corrconcept}. In the case of error-free measurements (top two panels), the contamination will simply be the fractions of off-diagonal terms of the correlation plot between the reference tracer and the  comparison tracer.  However, measurement uncertainties complicate the picture as they randomly scatter points into the off-diagonal parts of the plot, even in the case of perfect tracer, as illustrated in the bottom panel of the figure.

We denote $f_i$ as the probability that a given tracer measurement $t_i \pm \delta t_i$ is below the cut-off value $t_\mathrm{cut}$ (and therefore the target is classified as $b$). Then, $f_i$ is  expressed (for normally distributed errors) as:
\begin{equation}
    \label{eq:fi}
    f_i = \mathcal{P}\left(t_i < t_\mathrm{cut}\right) = \int_{-\infty}^{t_\mathrm{cut}} \prob{\hat{t}_i}{t_i} \, \textrm{d}\hat{t}_i  = \int_{-\infty}^{t_\mathrm{cut}} \mathcal{N}(\hat{t}_i; t_i, \delta t_i)\,\textrm{d}\hat{t}_i
\end{equation}
where $\hat{t}_i$ correspond to the true value of $t_i$.

By definition, one then has $N^b \equiv \sum f_i$ and $N^a \equiv \sum (1-f_i)$; similarly, the number of off-diagonal elements are $N_a^b = \sum_{i\in a} f_i$ and $N_b^a = \sum_{i\in b} (1-f_i)$.  Therefore, even in the context of a perfect tracer (for which one expects $c_a = c_b = c^a = c^b = 0$), the previous simple contamination estimates $c^a = N_b^a / N^a$ or $c^b = N_a^b / N^b$ can appear to be non-zero due to measurement uncertainties, and degrade the intrinsic tracer contamination estimates.

Rather, $c^a$ and $c^b$ should be estimated as the fractions of off-diagonal terms that are not caused by measurement errors; this will be done by defining their probability function, and comparing it to observations.

\subsection{Building the probability function} to estimate \texorpdfstring{$c^a$}{ca} and \texorpdfstring{$c^b$}{cb}
\label{sec:measuringcyco}
In order to get the contamination of a tracer ($c=c^a+c^b$; see Section~\ref{sec:notations}), we first express the probability function of the intrinsic parameters $c_a$ and $c_b$.

Following the derivations presented in Appendix~\ref{app:mathematical_derivation}, one can express the probability of measuring $t_i$ when the target $i$ belongs to population $b$ as:
\begin{equation}
    \mathcal{P}(i \in b, t_i \mid c_b) \propto  (1-p_a)\,\Big((1-c_b) \times f_i + c_b \times (1 - f_i)\Big).
\end{equation}
It can be understood as the probability that a target is $b$, i.e. $(1-p_a)$, times the chances that a tracer is measured below a cut ($f_i$) --~and thus classifying the target as "$b$"~-- while accounting for the fraction of false-negative ($1-c_b$), plus the chances that the tracer is measured above the cut ($1-f_i$) times the fraction of false-positive ($c_b$).

Similarly for the probability of measuring $t_i$ when the target $i$ belongs to population $a$:
\begin{equation}
    \mathcal{P}(i \in a, t_i \mid c_a) \propto  p_a\,\Big(c_a \times f_i + (1-c_a) \times (1 - f_i)\Big),
\end{equation}
and therefore:
\begin{align}
    \label{eq:likelihood_base}
    \mathcal{P}(t_i \mid c_a, c_b)
    &= \mathcal{P}(i \in a, t_i \mid c_a) + \mathcal{P}(i \in b, t_i \mid c_b)\nonumber\\
    &\propto p_a \times \Big( (1-c_a) \times (1-f_i) + c_a \times f_i \Big) \nonumber\\
    &\quad + (1-p_a) \times \Big(c_b \times (1-f_i) + (1-c_b)\times f_i \Big).
\end{align}

As we said in Section~\ref{sec:ref_tracer}, in reality, one does not know if an individual target belongs to the $a$ or $b$ class: the contamination can not be directly tied to the ``truth'', but only to another tracer used as a reference, plagued by its own contamination and measurement errors. Using ${}^\textrm{ref}$ to denote the parameters for the reference tracer, $p_a$ and $p_b=1-p_a$, which cannot be estimated directly anymore, can be derived from the reference tracer as:
\begin{align}
    p_a &\equiv \prob{i \in a, t_i^\textrm{ref}}{c_a^\textrm{ref}} \nonumber\\
    &= p_a^\textrm{ref} \times \left( (1-c_a^\textrm{ref}) \times (1-f_i^\textrm{ref}) + c_a^\textrm{ref} \times f_i^\textrm{ref} \right)
\end{align}
and
\begin{align}
    p_b &\equiv \prob{i \in b, t_i^\textrm{ref}}{c_b^\textrm{ref}} \nonumber\\
    &= (1-p_a^\textrm{ref}) \times \left(c_b^\textrm{ref} \times (1-f_i^\textrm{ref}) + (1-c_b^\textrm{ref})\times f_i^\textrm{ref} \right).
\end{align}
where $p_a^\textrm{ref}$, the fraction of truly $a$ targets, have to be assumed \emph{a priori} since the reference tracer is noisy (see Section~\ref{sec:cont_from_lssfr}).

Assuming the reference tracer to be perfect, i.e. $c_a^\textrm{ref}~=~c_b^\textrm{ref}~=~0$, Eq.~\ref{eq:likelihood_base} becomes:
\begin{align}
    \label{eq:probatracercomplete}
    \mathcal{P}(t_i, t_i^\textrm{ref} \mid c_a, c_b) \propto \qquad& \nonumber\\
    p_a^\textrm{ref} \times (1-f_i^\textrm{ref}) &\times \bigg( (1-c_a) \times (1-f_i) + c_a \times f_i \bigg) \nonumber\\
    {} + (1-p_a^\textrm{ref}) \times f_i^\textrm{ref} &\times \bigg(c_b \times (1-f_i) + (1-c_b)\times f_i \bigg).
\end{align}

Finally, the estimation of the tracer's parameters $c_a$ and $c_b$ with respect to the reference tracer is made by minimizing:
\begin{equation}
    \label{eq:likelihood}
    \mathcal{L} = -2 \sum_i \ln \mathcal{P}(t_i, t_i^\textrm{ref} \mid c_a, c_b).
\end{equation}
and the total contamination $c = c^a + c^b$ for a comparison tracer can be computed from:
\begin{align}
    c^a &= \frac{N_b}{N^a} c_b = \frac{(1-p_a^\textrm{ref})N}{\sum (1-f_i)} c_b \nonumber\\
    c^b &= \frac{N_a}{N^b} c_a = \frac{p_a^\textrm{ref} N}{\sum f_i} c_a.
\end{align}

We have tested and validated our model and our code using simulations. We generated mock dataset of various contaminations and sizes, which have then been fitted with our implementation of the likelihood described in this section. The results confirm that our implementation of the algorithm is correct.

\section{Data}
\label{sec:data}

We work with the Nearby Supernova Factory \citep[SNfactory, ][]{aldering2002} SNe~Ia dataset published in \citet[see also the \citealt{aldering2020}]{rigault2020}. This dataset has two benefits for this analysis: (1) it is at low-redshifts ($0.02<z<0.08$), so the local environment is measurable, and (2) it contains spectrophotometric IFU environmental data, necessary to accurately estimate the local specific Star Formation Rate (lsSFR). We use the publicly available catalog and images from SDSS and PS1 for the photometric measuments or their derived quantities. 

This section briefly summarizes the methodology to extract the different tracers considered in this analysis, which follows those developed in the literature,  namely: the spectroscopically-derived lsSFR (Section~\ref{sec:data_LsSFR}), photometrically-derived lsSFR (Section~\ref{sec:photolssfr}), local colors (Section~\ref{sec:datacol}), local and global host stellar masses (Section~\ref{sec:datamass}) and the global host morphologies (Section~\ref{sec:data_morpho}).

\subsection{Spectroscopic lsSFR}
\label{sec:data_LsSFR}

The spectroscopically-derived lsSFR is detailed in Section~3 of \cite{rigault2020}. We use their measurements, which are generated in two stages: first, the Star Formation Rate (SFR) is derived from the H$\alpha$ emission line luminosity \citep{calzetti2013}, spectroscopically measured within the local 1~kpc aperture radius, after subtraction of the stellar continuum background. The second step is the measurement of the local stellar mass, as later described in Section~\ref{sec:datamass}. For both quantities, a full posterior distribution is derived such that their ratio sets the posterior distribution of the lsSFR measurements. Hereafter, we will refer to this tracer as the ``spectroscopic lsSFR''.
 
\subsection{Photometric measurements}
\label{sec:photodata}

We use flux-calibrated optical images from SDSS \citep[DR12,][]{alam2015} to derive the photometric environmental tracers.

We measure $ugriz$ SDSS local fluxes and their uncertainties in projected circular apertures centered on the SN location, using the \textsc{sum\_circle} method of \sep{}\footnote{\href{http://github.com/kbarbary/sep}{github.com/kbarbary/sep} v1.10} \citep{barbary2016}. To compare our results with literature studies, we use aperture radii of $X = 1$, $1.5$ and $3$~kpc. Counts are converted to flux assuming a zero point of $22.5$~mag for the $gri$ bands and 22.46 and 22.52 for the $u$ and $z$ bands, respectively.

To test the accuracy of the sky background subtraction, we draw 500 random source-free apertures around each target. Presumably, the histogram of the error-normalized background levels should be a standard $\mathcal{N}(0, 1)$ pull distribution. However, we regularly observe that the pull mean is slightly too high, corresponding to an inaccurate background correction, and that the pull dispersion is larger than unity, meaning that the error on the background level has been under-estimated.  We thus further correct each aperture photometric measurement by the median of the 500 random ``sky'' apertures, and scale the quoted error by the normalized median absolute deviation of the sky levels.

To derive global tracers, we first associate the SN with its host employing the Directional Light Radius method \citep{sullivan2006,gupta2016}; ellipses used to determine the directional light radius are obtained by \textsc{sum\_ellipse} method of \sep{}. Then we use the $ugriz$ global galaxy model magnitudes and fluxes from the corresponding SDSS catalog entries (see details in \citealt{rigault2020}).

Both local and global photometric measurements are then corrected for Milky Way dust absorption using the \pkg{extinction}\footnote{\href{http://extinction.readthedocs.io/}{extinction.readthedocs.io}} library assuming a \cite{fitzpatrick1999} extinction curve with $R_V=3.1$ and the dust extinction map from \cite{sfd1998}.

\subsection{SED fitting and k-correction}
\label{sec:sedfit}

From each $ugriz$ photometric dataset, we use \pkg{LePhare}\footnote{v2.2 see \href{http://www.cfht.hawaii.edu/\~arnouts/LEPHARE/lephare.html}{LePhare website}} \citep{arnouts1999,ilbert2006,arnouts2011} to fit for the associated Spectral Energy Distribution (SED) using  \cite{bruzual2003} templates (hereafter BC03) as did, e.g., \cite{jones2018}. 
For the reproducibility, our configuration file is available online\footnote{\href{https://github.com/MartinBriday/pylephare}{github.com/MartinBriday/pylephare}}. 
It contains the following assumptions:

\begin{description}
    \item[dust:] we use 41 bins of $E(B-V)$ extinction values ranging from 0 to 1 (per 0.01 step from 0 to 0.2, 0.03 from 0.2 to 0.5 and then a 0.05 step up to 1) and we use the \cite{fitzpatrick1999} extinction curve \citep[extracted from the hyper-$z$ program with $R_V=3.1$; ][]{bolzonella2000};
    \item[redshift range:] we use a redshift range from 0 to 0.1 with a bin size of $\Delta z=0.002$;
    \item[emission lines:] this contribution is included \citep{kennicutt1998}; 
    \item[cosmology:] we use $H_0=70\;\mathrm{km\,s^{-1}\,Mpc^{-1}}$, $\Omega_m=0.3$, and  $\Omega_{\Lambda}=0.7$.
\end{description}

The SED fit is made at the fixed (known) redshift of the host, the stellar mass $\mathrm{M_*}$ is bounded between $10^6$ and $10^{13}$ solar masses, and the $r$-band absolute magnitude between $-10$ and $-26$.
We include SDSS' suggested error floor ($0.05, 0.02, 0.02, 0.02, 0.03$ for the $ugriz$ bands, respectively, see e.g., \href{http://kcorrect.org/}{kcorrect.org} and \citealt{childress2013a}).

We estimate the posterior distribution of each SED fitted parameter and spectra using Monte Carlo simulations. For each $ugriz$ flux measurement, i.e. for each local radius of each SN, we randomly draw 500 realisations assuming the bands are independent and that flux errors are normally distributed. We run the SED fitting procedure for these 500 realisations and the best fitted parameters (\pkg{\_best}) and spectral distributions set the respective posteriors. 
We use the median rest-frame magnitudes measured on each of the 500 realisation to set the $ugriz$ k-corrected magnitudes used in this analysis; the 16\% and 84\% percentiles set the corresponding errors.

\subsection{Colors}
\label{sec:datacol}
Colors are estimated from the k-corrected magnitudes (see Section~\ref{sec:sedfit}): the $u-r$ color is the difference of the median of the k-corrected $u-$ and $r-$band magnitudes and the color error is the quadratic sum of the individual standard deviations.

\subsection{Stellar Masses}
\label{sec:datamass}

Local and global masses are derived using the procedure described in Section~3.3 of \cite{rigault2020}; see also \cite{jones2018}. In brief, we use the relation from \cite{taylor2011} to convert $g$ and $i$ k-corrected magnitudes into stellar masses.
This relation has a 0.1~dex intrinsic dispersion that is added in quadrature to the stellar mass uncertainties derived from photometric uncertainties only; this scatter dominates the error budget, especially for global measurements.
See \cite{smith2020} for a discussion about the consistency of stellar mass estimators in the context of SN host analyses. 

\subsection{Photometric lsSFR}
\label{sec:photolssfr}

\cite{jones2018} use photometry-based sSFR estimation to assess the lsSFR parameter in place of the H$\alpha$-based measurements, as they do not have local spectroscopy. They employ \pkg{LePhare} in a similar fashion as described in Section~\ref{sec:sedfit} and estimate their sSFR posterior (and its errors) from the 50\% ([16\%, 84\%]) of the individual sSFR values from the Monte Carlo realisations. They use the SDSS \textit{u-}band plus $grizy$ from PanStarrs DR1 \citep[PS1,][]{chambers2016} to do so. To be consistent, when deriving the sSFR, we also use these data, applying the same photometric measurement on PS1 data as we did for SDSS (see Section~\ref{sec:photodata}). Calibrated PS1 images are downloaded from the cutout service\footnote{\href{https://outerspace.stsci.edu/display/PANSTARRS/PS1+Image+Cutout+Service}{PanSTARRS website.}}. Still for consistency, we also use \cite{jones2018}'s \pkg{LePhare} configuration file when measuring the sSFR this way (D. Jones private communication). 
The measurements are finally normalized by the surface area. Hereafter we refer to this tracer as the ``photometric lsSFR'', in contrast to the spectroscopic one.

\subsection{Morphology}
\label{sec:data_morpho}

The inverse concentration index \citep[\ici,][see also \href{https://www.sdss.org/dr12/algorithms/classify/}{SDSS web site}]{shimasaku2001, strateva2001}
is a commonly used morphological tracer. 
It is the ratio between radii containing 50\% and 90\% of the Petrosian flux in $r$-band.  
With this tracer, early-type galaxies typically have an $\ici\sim0.3$, while late-type galaxies have an \ici{} closer to 0.45. Following \cite{kauffmann2003}, we use $\ici_\mathrm{cut}=1/2.7=0.37$ to distinguish between late and early-type galaxies \citep[see also, e.g.,][]{choi2010}. Using slightly lower boundaries such as 0.35 suggested by, e.g., \cite{banerji2010} has marginal influence on our results.

Galaxy classification based on the Petrosian flux is one approach, but not unique. Among a few others, the \ici{} is the most convenient for this analysis as it discriminates two galaxy morphology populations (early/late types), separated by a cut-off value, motivating our choice.

\section{Contamination with respect to the spectroscopic lsSFR}
\label{sec:cont_from_lssfr}

Because the sSFR is usually used as a reference age tracer when available \citep[see e.g.][]{yoshikawa2010, labbe2013, casado2015, karman2017}, we use the spectroscopic lsSFR measurement from \cite{rigault2020} as a reference tracer, i.e., we assume that this quantity is a perfectly accurate (yet imprecise) progrenitor age tracer: $c^a_{\mathrm{spec-lsSFR}}~=~c^b_{\mathrm{spec-lsSFR}}~=~0$; see Section~\ref{sec:concept}. We test this hypothesis in Section~\ref{sec:reference_tracer}.

Following most preceding host environmental studies, which split their samples at the median value, we will also assume throughout the paper that $p_a^\textrm{ref} = 50\%$. We have found by simulations that the derivation of the $c_a$ and $c_b$ tracer parameters are unaffected by this choice (bias lower than the $1\sigma$ error), as long as the actual true parameter $25\% \lesssim p_a^\textrm{ref} \lesssim 75\%$, which is to be expected simply based on rate analyses \citep[e.g.][]{mannucci2006, rodney2014, wiseman2020}; bias becomes significant (at more than $3\sigma$) if $p_a<10\%$ or $p_a>90\%$.

\subsection{Comparing the environmental tracers with the spectroscopic lsSFR.}
\label{sec:comparison_tracers}

Following \cite{rigault2020}, we classify as population $a$ every SNe~Ia with $\log(\textrm{spec-lsSFR})> -10.82$~dex, as this value corresponds to the median of this tracer for the SNfactory sample. Given the measurement errors, each SN~Ia therefore has a probability $p(young)$ to be observed in $a$ population (referring to $1-f_i^\textrm{ref}$ in Section~\ref{sec:measuringcyco}). As highlighted in Section~\ref{sec:concepterr}, the sum of the off-diagonal terms are fully captured by the measurement errors since we assumed $c^a_{\mathrm{spec-lsSFR}}=c^b_{\mathrm{spec-lsSFR}}=0$.

Each tracer has its own cut-off value to classify as $a$ a given SN~Ia. We will apply the assumption typically used in the literature when available and the median value otherwise, which is usually also the literature assumption. The well studied mass step usually has the threshold boundary set at $\log({M_*}/{M_{\odot}})_\mathrm{cut} = 10$~dex \citep[for e.g.][]{kelly2010,sullivan2010, betoule2014, scolnic2018} and we will use that value in this analysis. Note that since we have $\sim60\%$ of SNe~Ia with a host-mass greater than $10^{10}\;M_{\odot}$ \citep[in agreement with e.g.,][]{roman2018}, and since $N_a/N$ is given by the reference tracer (here $p_a^\textrm{ref} = 50\%$), this means that $c^a_{\mathrm{gmass}}\neq c^b_{\mathrm{gmass}}$. Concerning the local mass, \cite{jones2018} and \cite{rigault2020} use the median to divide their respective samples. For the sample here, we find the median to be $\log({M_*}/{M_{\odot}})_\mathrm{cut} = 8.37$~dex; to compare with \cite{jones2018}, we use a $1.5$~kpc ``local'' aperture for the local mass. Following \cite{roman2018}, \cite{jones2018} and \cite{kelsey2021}, we do the same for the local ($3$~kpc) color and the photometric local ($1.5$~kpc) sSFR finding medians at $u-r=1.74$~mag and $\log(\textrm{phot-lsSFR})=-10.32$~dex, respectively. Finally, as explained in Section~\ref{sec:data_morpho}, we use $\ici = 0.37$ to divide the morphologies of our SN Ia host galaxies into category $a$ or $b$, respectively above or below this value.

\begin{figure*}
    \centering
    \includegraphics[width=0.65\linewidth]{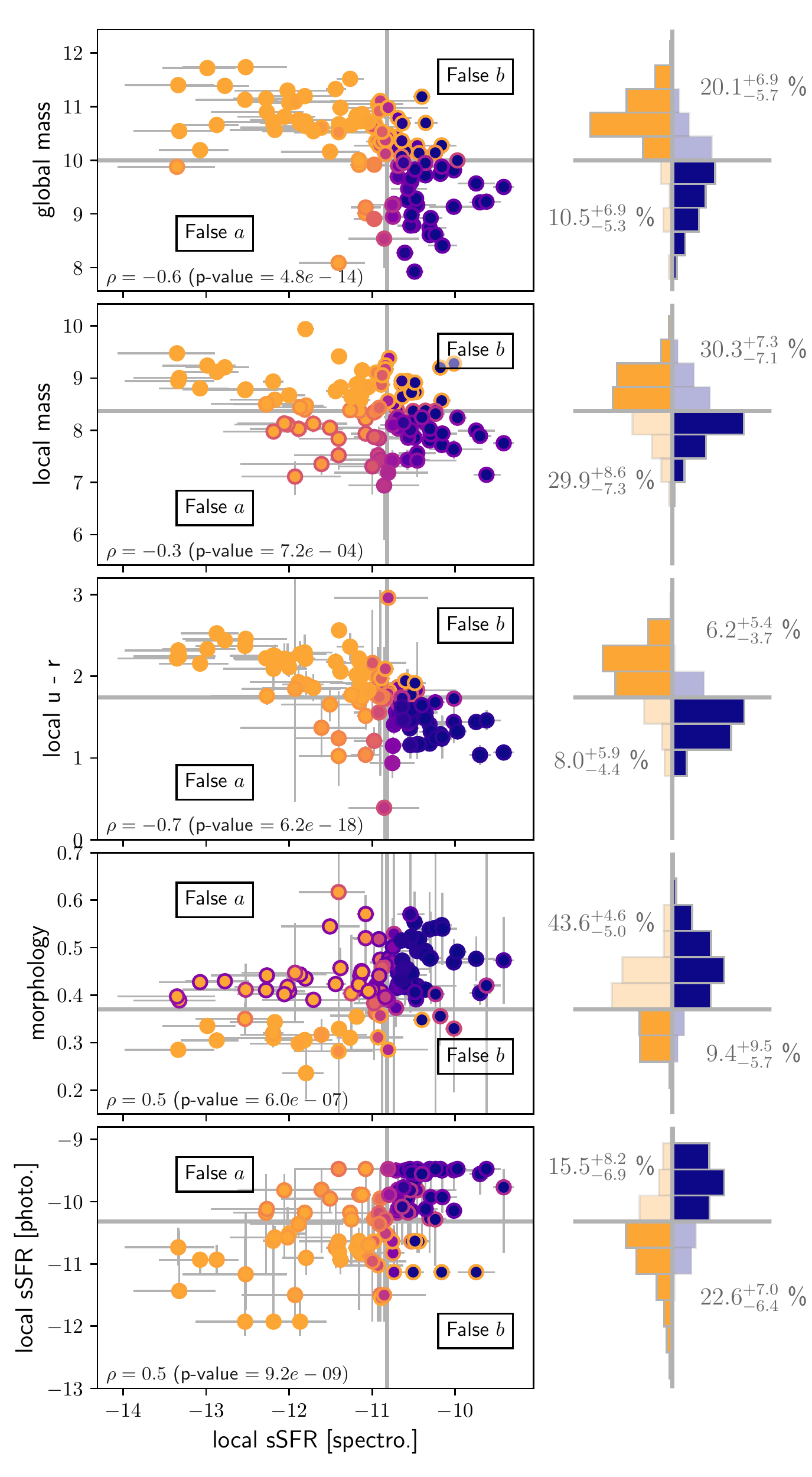}
    \caption{Correlation between each environmental tracer and the spectroscopic lsSFR, as mentioned in Section~\ref{sec:cont_from_lssfr}. The vertical grey line is the spectroscopic lsSFR cut-off, set at $\log(\textrm{lsSFR}) = -10.82$~dex. The horizontal grey lines are the other tracer cut-offs, as defined in Section~\ref{sec:cont_from_lssfr}. Each figure indicates the false $a$ and false $b$ classification quadrants. The point face-color, varying from orange to blue, represents the probability for a SN~Ia to be in $a$ (blue) or $b$ (orange) populations in the spectroscopic lsSFR point of view, whereas the point edge-color represents the same probability but from the comparison tracer point of view. In the right column, the histograms plot the distributions in each quadrant. Orange (resp. blue) bars correspond to the truly $b$ (resp. $a$) spectroscopic lsSFR's classification. The estimated $c^a$ and $c^b$ parameters are printed in percentages within the corresponding quadrants where the histogram is transparent.}
    \label{fig:tracer_corr}
\end{figure*}

In Fig.~\ref{fig:tracer_corr}, we show the correlation between the spectroscopic lsSFR and the other environmental tracers. We first note that every tracers correlate relatively well with the spectroscopic lsSFR. Based on the Spearman rank coefficient, the most correlated tracer is the local $u-r$ color ($|\rho|=0.71$) closely followed by the global mass ($|\rho|=0.64$); the host galaxy morphology and the local stellar mass show weaker correlations, with $|\rho|=0.45$ and $0.32$, respectively. 

The population of off-diagonal terms appear to be consistent with these Spearman-ranked correlations: the higher the fraction of off-diagonal terms, the lower the Spearman coefficient value. Yet, as detailed in Section~\ref{sec:concepterr}, only the fraction of off-diagonal terms not caused by measurement errors has to be accounted for to measure the accuracy of an indicator to trace the reference, here the spectroscopic lsSFR.

To do that, we fit the tracers' $c_a$ and $c_b$, assuming that the spectroscopic lsSFR measurements are (noisy) perfectly accurate indicators of $a$ and $b$, by minimizing Eq.~\ref{eq:likelihood}. Since the measurements are noisy, we need to set the fraction of truly $a$, that we assume to be 50\%, i.e. we fix $N_a=N_b$. The fit is made using Markov Chain Monte Carlo using the \pkg{emcee} \citep{foreman2013} package to sample the full posterior distributions of the tracers' $c_a$ and $c_b$ parameters. Each tracer is fitted independently. The resulting median parameters $c^a$ and $c^b$ are displayed in Fig.~\ref{fig:tracer_corr} with their $1\sigma$ scale (16\%/84\%), and the aggregated tracer contamination $c=c^a+c^b$ is summarized in Table~\ref{tab:results}.

\subsection{Measuring the environmental magnitude step}
\label{sec:standardization}

To derive the SNe~Ia magnitude steps, we follow the procedure detailed in Section~4.2.2 of \cite{rigault2020}. Given $p_i^t=1-f^t_i$ the probability that the tracer measurement is above the tracer's cut-off value (see Section~\ref{sec:concepterr}), thus classified as $a$ by this tracer, we fit the magnitude offset $\gamma$ between the two populations (aka, the step) together with the stretch and color standardisation coefficients $\alpha$ and $\beta$. This is done by $\chi^2$ minimisation between $\mu_{\mathrm{\Lambda CDM}}$ and the standardized SN~Ia distance modulus:
\begin{equation}
    \mu = m - M + \alpha x_1 - \beta c + \gamma p^t.
    \label{eq:salt_standardization}
\end{equation}

When doing the fit, we fix the cosmology \citep{planck2015} and the covariances between $m$, $x_1$ and $c$ are taken into account; $p^t$ is an independent measurement.

When fitting each environmental tracer independently to derive its associated step $\gamma$, the $\alpha$ and $\beta$ coefficients are free to vary and might therefore differ between tracers. However, if all tracers are probing the same underlying effect, $\alpha$ and $\beta$ should be the same, since the stretch and color standardisation should not depend on the accuracy with which one is able to probe this underlying effect. When fitting $\alpha$, $\beta$ and $\gamma$ simultaneously, because stretch (and color) are connected to the host properties, the recovered value of $\alpha$ and $\beta$ will be unbiased if the true underlying tracer is used to determine $\gamma$ \citep{dixon2021}.
We further investigate this issue in Section~\ref{sec:alphabeta_of_lssfr}, where the $\alpha$ and $\beta$ will be fixed to those derived together with the step of the reference tracer.

\section{Results}
\label{sec:results}

\begin{figure*}
    \centering
    \includegraphics[width=\linewidth]{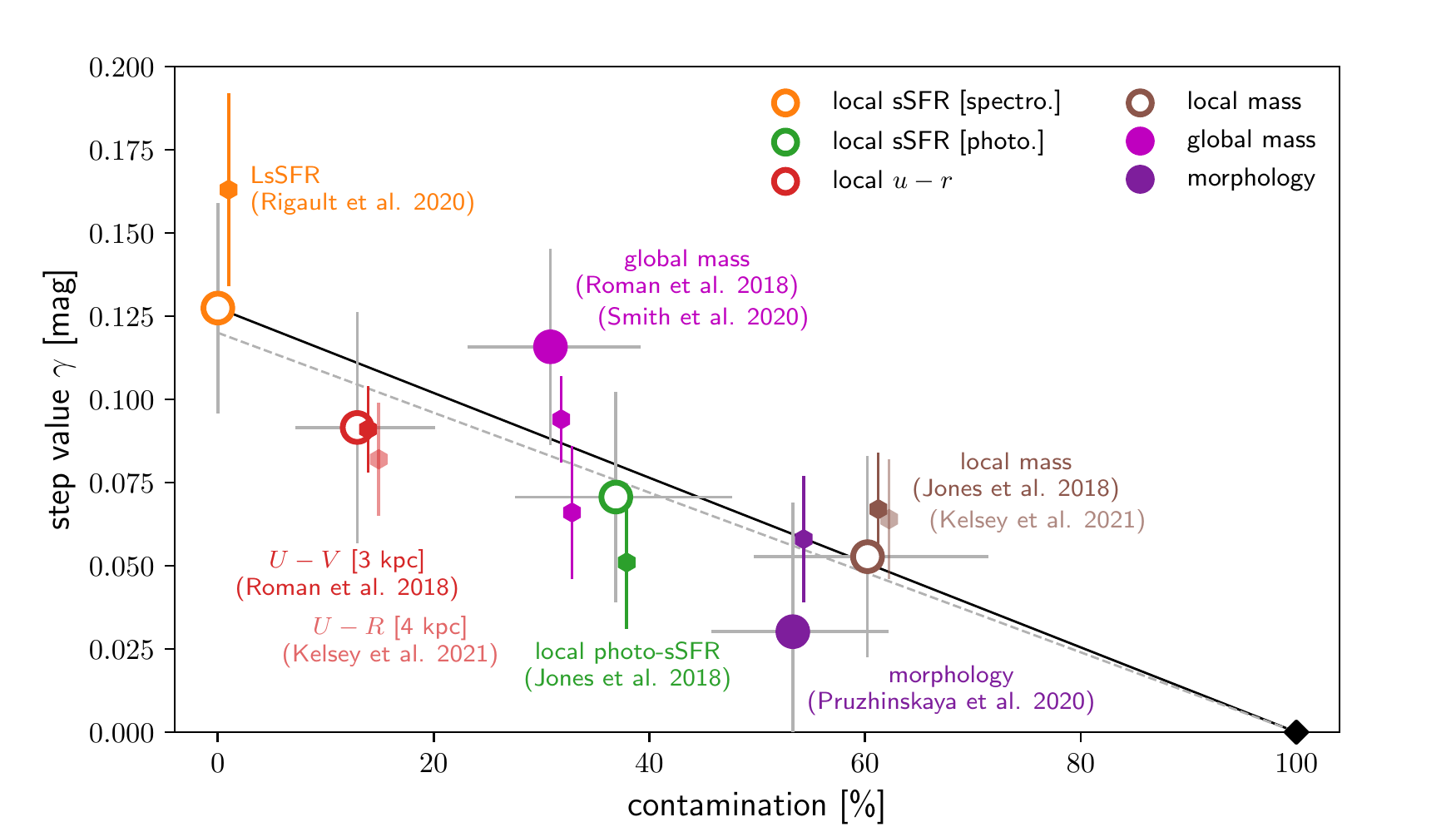}
    \caption{SNe~Ia magnitude steps, $\gamma$, as a function of the tracer contamination $(c=c^a+c^b)$, using the spectroscopic lsSFR as reference ($c_\mathrm{spec-lsSFR}\equiv 0$). Large markers are the steps derived using the SNfactory dataset (this work) while small hexagons are literature results (see Section~\ref{sec:results}). Large open (resp. full) circles are local (resp. global) measurements (see legend); the SNfactory-based and literature results have the same colors. The literature contaminations are those derived using the SNfactory sample and have been shifted by $+1\%$ for visibility. The \cite{kelsey2021} related hexagonal markers are transparent as they measure the magnitude step after standardization, while all other datapoints (both with our SNfactory sample and literature results) fit $\gamma$ as a third standardization parameter (see Section~\ref{sec:standardization}). The full black diamond indicates the 0 mag step associated by definition to 100\% contamination, corresponding to a random SN population classification. The straight black line shows our model of the measured step as a function of the contamination, linking the reference tracer point to the black diamond (see Section~\ref{sec:concept}). The dashed grey line is a fit to the literature measurements, constrained to pass through the random classification (black diamond) value.}
    \label{fig:envdiaglssfr}
\end{figure*}

The derived tracer contaminations with respect to the spectroscopic lsSFR and their associated magnitude steps $\gamma$, using the SNfactory dataset, are summarized in Table~\ref{tab:results} and shown in Fig.~\ref{fig:envdiaglssfr}. 

\begin{table*}
    \centering
    \caption{Table comparing the standardisation coefficients $\alpha$, $\beta$ and $\gamma$ (aka. the stretch and color coefficients and the magnitude step value) and the tracer intrinsic contamination ($c=c^a+c^b$) with respect to the spectroscopic lsSFR used as a reference tracer; $\gamma$* is the magnitude step value when fixing $\alpha$ and $\beta$ to those obtained by the reference tracer standardisation (see Section~\ref{sec:alphabeta_of_lssfr}); $t_{cut}$ is tracer's cut-off.}
    \label{tab:results}
    \renewcommand{\arraystretch}{1.2}
    \begin{filecontents*}{table_results.csv}
tracer,cut,contint,alph,bet,magstepsample,magstepsamplefixed,magsteplitt
local sSFR [spectro.],${-10.82}$,${0.0}$,${-0.156} \pm {0.013}$,${3.042} \pm {0.128}$,${0.127} \pm {0.032}$,${0.127} \pm {0.032}$,${0.163} \pm {0.029}$
local sSFR [photo.],${-10.32}$,${36.9}_{-9.3}^{+10.8}$,${-0.138} \pm {0.012}$,${3.059} \pm {0.128}$,${0.071} \pm {0.032}$,${0.078} \pm {0.032}$,${0.051} \pm {0.020}$
local $u - r$,${1.74}$,${12.9}_{-5.7}^{+7.2}$,${-0.150} \pm {0.013}$,${3.115} \pm {0.132}$,${0.092} \pm {0.035}$,${0.095} \pm {0.030}$,${0.091} \pm {0.013}$
local mass,${8.37}$,${60.2}_{-10.6}^{+11.2}$,${-0.137} \pm {0.012}$,${3.142} \pm {0.138}$,${0.053} \pm {0.030}$,${0.051} \pm {0.028}$,${0.067} \pm {0.017}$
global mass,${10.00}$,${30.8}_{-7.7}^{+8.3}$,${-0.151} \pm {0.013}$,${3.121} \pm {0.132}$,${0.116} \pm {0.030}$,${0.116} \pm {0.027}$,${0.094} \pm {0.013}$
morphology,${0.37}$,${53.3}_{-7.5}^{+8.9}$,${-0.138} \pm {0.013}$,${3.044} \pm {0.139}$,${0.030} \pm {0.039}$,${0.051} \pm {0.036}$,${0.058} \pm {0.019}$
\end{filecontents*}
\csvreader[tabular=l c c c c c c, table head=\hline \hline 
    Tracer & $t_{cut}$ & $c^a+c^b$ & $\alpha$ & $\beta$ & $\gamma$ & $\gamma$*
    \\\hline, table foot=\hline, head to column names]
    {table_results.csv}
    {}
    {\tracer & \cut & \contint & \alph & \bet & \magstepsample & \magstepsamplefixed}
\end{table*}

The step amplitudes as a function of the tracer contaminations follow the expected trend given by Eq.~\ref{eq:stepbias}, shown by the straight black line in the figure, remarkably well. As detailed in Section~\ref{sec:concept}, this diagonal --~going from $\gamma_0$ at $c=0\%$ (the reference tracer step value) to 0 at $c=100\%$ (the black diamond in the figure)~-- is expected if: (1) two SN~Ia populations exist with different mean magnitudes, and (2) we are using tracers that are not perfectly able to discriminate the two populations ($c>0$) to measure their magnitude offset. We emphasize that $c=100\%$ corresponds to randomly distributed SNe~Ia between the two classes, the expected magnitude difference between the two resulting groups thus is 0 by definition (as seen in Fig.~\ref{fig:contconcept}).

We added to Fig.~\ref{fig:envdiaglssfr} recent results from the literature with the step measurements that were made using the same techniques as the ones we used. Namely, we added the lsSFR from \cite{rigault2020}, the local ($3$~kpc) $U-V$ (similar to $u-r$) and global host stellar mass (split at $10^{10}\,M_{\odot}$) from \cite{roman2018}, the local ($1.5$~kpc) stellar mass and photometric sSFR from \cite{jones2018} and the morphology from \cite{pruzhinskaya2020}. We used the global host mass-step from \cite{roman2018} for it is derived using the state-of-the-art \cite{betoule2014}~+~SNLS-5 years sample and the Malmquist bias correction were not made using the ``5D'' implementation of the Beams with Bias Correction \citep[BBC;][]{scolnic2016, kessler2017}. \citet{smith2020} showed that this implementation can bias the reported step if intrinsic SN-host correlations are not accounted for. We also used the global host mass step from \cite{smith2020}. Finally, we plotted in this figure the local ($4$~kpc) $U-R$ and stellar mass from \cite{kelsey2021} with a transparent marker, as $\gamma$ is fitted after the standardization in that paper (while we fit it as a third standardization parameter, see Section~\ref{sec:standardization}).
For these literature datapoints, we use the reported steps while using our derived tracer contaminations. We highlight that, if one considers the steps to be SN sample dependent (e.g., due to the lightcurve extraction pipeline), that implies that the contaminations are purely galaxy properties that are unrelated to the SNe~Ia.

The Fig.~\ref{fig:envdiaglssfr} shows that step amplitudes measured using the SNfactory data, measured using any of the literature environmental tracers, are in remarkable agreement with the corresponding independent literature measurements. For instance \cite{jones2018} local mass step is $0.067 \pm 0.017$~mag, while we measure $0.053 \pm 0.031$~mag and the local $U-V$ color step from \cite{roman2018} is $0.091 \pm 0.013$~mag and we find $0.096 \pm 0.035$~mag. The SNfactory SNe~Ia data thus seem to be representative of that of the literature.

Under the assumption of multiple populations, implicitly implied by the environmental step functional form, the fact that the relationship between observed environmental step and tracer contamination is compatible with our two-populations model, for both SNfactory and the literature datapoints, suggests: (1) that two populations are enough to explain the observations; and (2) that both populations differ in standardized brightness by $\gamma_0\sim0.13$~mag. Fitting for $\gamma_0$ using literature data points, as shown in Fig.~\ref{fig:envdiaglssfr}, we find $0.121\pm0.010\,\mathrm{mag}$. This claim assumes that spectroscopic lsSFR is a perfect tracer. We study the use of the other tracers as the reference tracer in Section~\ref{sec:reference_tracer}.

In reality, no tracer is perfect and if the spectroscopic lsSFR contamination were to be, say, a few percents, then $\gamma_0$ would actually be higher. For instance if $c_{\mathrm{spec-lsSFR}}=10\%$ and $\gamma_{\mathrm{spec-lsSFR}}=0.13$~mag then $\gamma_0 = \gamma_{\mathrm{spec-lsSFR}} (1-c)^{-1}=0.145$~mag. Consequently, the reported $\gamma_0$ measurements made in this analysis assuming we have a noisy but perfect tracer are, in fact, lower-limits on the actual SN~Ia population difference in magnitude means.
The true spectroscopic lsSFR contamination is beyond the scope of this work.

\section{Discussion}
\label{sec:discussion}

In this section, we present variations to the main analysis and then discuss the consequences of our findings. We first study the impact of fixing the stretch and color standardisation coefficients to that of the reference tracer. We then change which tracer is used as a reference and compare their ability to describe the data.

\subsection{Fixing \texorpdfstring{$\alpha$}{alpha} and \texorpdfstring{$\beta$}{beta}}
\label{sec:alphabeta_of_lssfr}

In Section~\ref{sec:results}, we fit the standardisation coefficients $\alpha$, $\beta$ and $\gamma$ for each environmental tracer independently to find that the two-populations model detailed in Section~\ref{sec:concept} seems to explain the observed variations between the tracer $\gamma$ parameters. Their apparent inconsistency is due to the ability of a tracer to accurately distinguish the underlying two populations. In that context, because the standardisation coefficients are correlated, especially $\alpha$ and $\gamma$ \citep[see e.g. Fig.~7 of ][]{rigault2020}, if one is not able to accurately measure $\gamma$ since its environmental tracer is inaccurate, one will in turn bias the derivation of the other standardisation parameters, as the fitter will use them to counter balance the $\gamma$ error.

The natural solution in the context of the two-populations model, which is implicitly assumed when doing a step analysis, is that one has to fix the value of $\alpha$ and $\beta$ to those derived when using the reference tracer when fitting Eq.~\ref{eq:salt_standardization} for the comparison tracers.

\begin{figure*}
    \centering
    \includegraphics[width=1\linewidth]{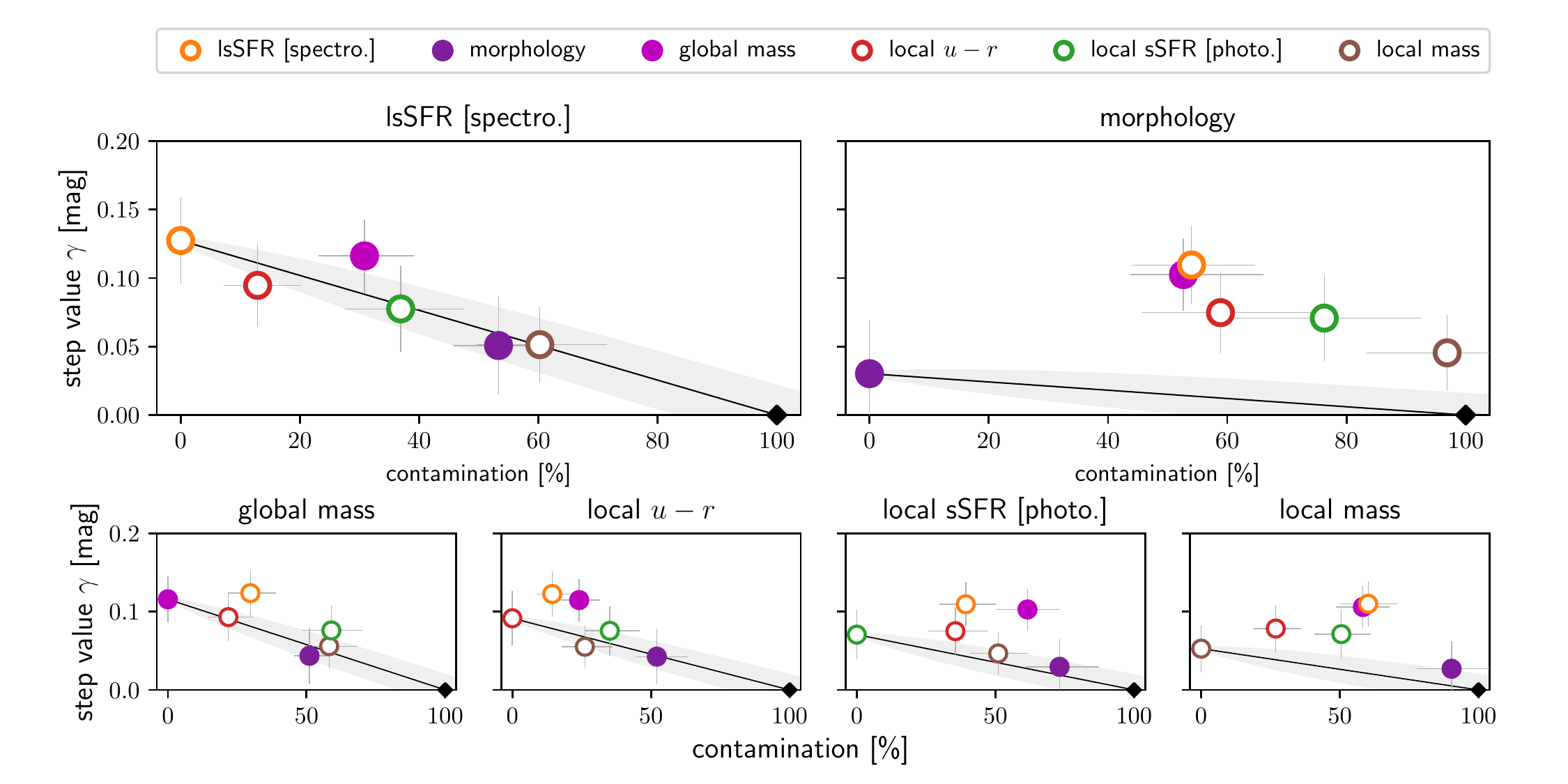}
	\caption{Similar to Fig.~\ref{fig:envdiaglssfr}, but changing the reference tracer choice by, from left to right, up to down, the spectroscopic lsSFR, the host galaxy morphology, the global host galaxy stellar mass, the local $u-r$ color, the photometric lsSFR and the the local stellar mass, respectively. In this figure, the $\alpha$ and $\beta$ have been fixed to that of the reference tracer used in each subplot (see Section~\ref{sec:alphabeta_of_lssfr}). The grey band around the black line is the expected scatter along the diagonal (see Section~\ref{sec:reference_tracer}). The more data deviate from the model expectation (black line), the less likely the reference tracer is closely connected to the actual underlying astrophysical origin.}
	\label{fig:tracer_envdiags}
\end{figure*}

The results of this alternative, yet more accurate, analysis is given in Table~\ref{tab:results} (column named $\gamma$*) and illustrated in the top-left plot in Fig.~\ref{fig:tracer_envdiags}. We see, comparing this plot with Fig.~\ref{fig:envdiaglssfr}, that the results converge on the model's expectations.

\subsection{Testing the reference tracer}
\label{sec:reference_tracer}

In this section, we vary which tracer is used as a reference tracer and we re-derive the resulting contamination terms, $c^a$ and $c^b$, as well as the steps $\gamma$* assuming the reference's $\alpha$ and $\beta$ as detailed Section~\ref{sec:alphabeta_of_lssfr}. If a tracer is a good reference, that is, if it is accurate at discriminating the true underlying two populations, then the other tracers should follow the diagonal line given by Eq.~\ref{eq:stepbias}, anchored at the value of $\gamma_0$ of the reference tracer. If a reference tracer is bad, the contamination associated to this tracer does not probe the ability of a comparison tracer to discriminate the underlying two populations. In that case, the points are not expected to follow the diagonal model.

This is what we qualitatively observe in Fig.~\ref{fig:tracer_envdiags}. Morphology is a bad reference tracer, as the other tracers lie far from its expected diagonal. This means that the morphology is not able to accurately discriminate the two underlying populations causing the environmental steps observed by the different tracers. Conversely, the spectroscopic lsSFR, the global mass and the local $u-r$ colors seem to be better reference tracers.

To quantify this observation, we first need to model how much scatter we expect along the diagonal if we did had access to a perfect tracer. This is mandatory since the step measurements are not independent, as they all are made from the same sample of SNe~Ia, but using different galaxy property indicators.

We use the simulation tool from Section~\ref{sec:concept} to simulate a sample with the same characteristic as the SNfactory one: $N=110$, $p_a=0.5$, $\sigma_a=\sigma_b=0.1$ and the $\gamma_0$ corresponding to the reference tracer step $\gamma$ in Table~\ref{tab:results}. 
We then assume a $c_a$ and a $c_b$, which define the four $N_i^j$ with $i=\{a,b\}$ and $j=\{a,b\}$. We randomly shuffle the sample to follow these $N_i^j$ and we measure $\gamma$. This last step is repeated 5000 times to determine the scatter on $\gamma$ caused by the randomness of which target belongs to the off-diagonal terms or not. If the two-populations model is correct, and if we measure the tracer $\gamma$ parameters with a single dataset, then this scatter corresponds to the expected variations given tracer $c_a$ and $c_b$ parameters. The amplitude of this scatter as a function of the tracers contaminations is shown as a grey band along the model's diagonal in Fig.~\ref{fig:tracer_envdiags}.

\begin{figure}
    \centering
    \includegraphics[width=1\linewidth]{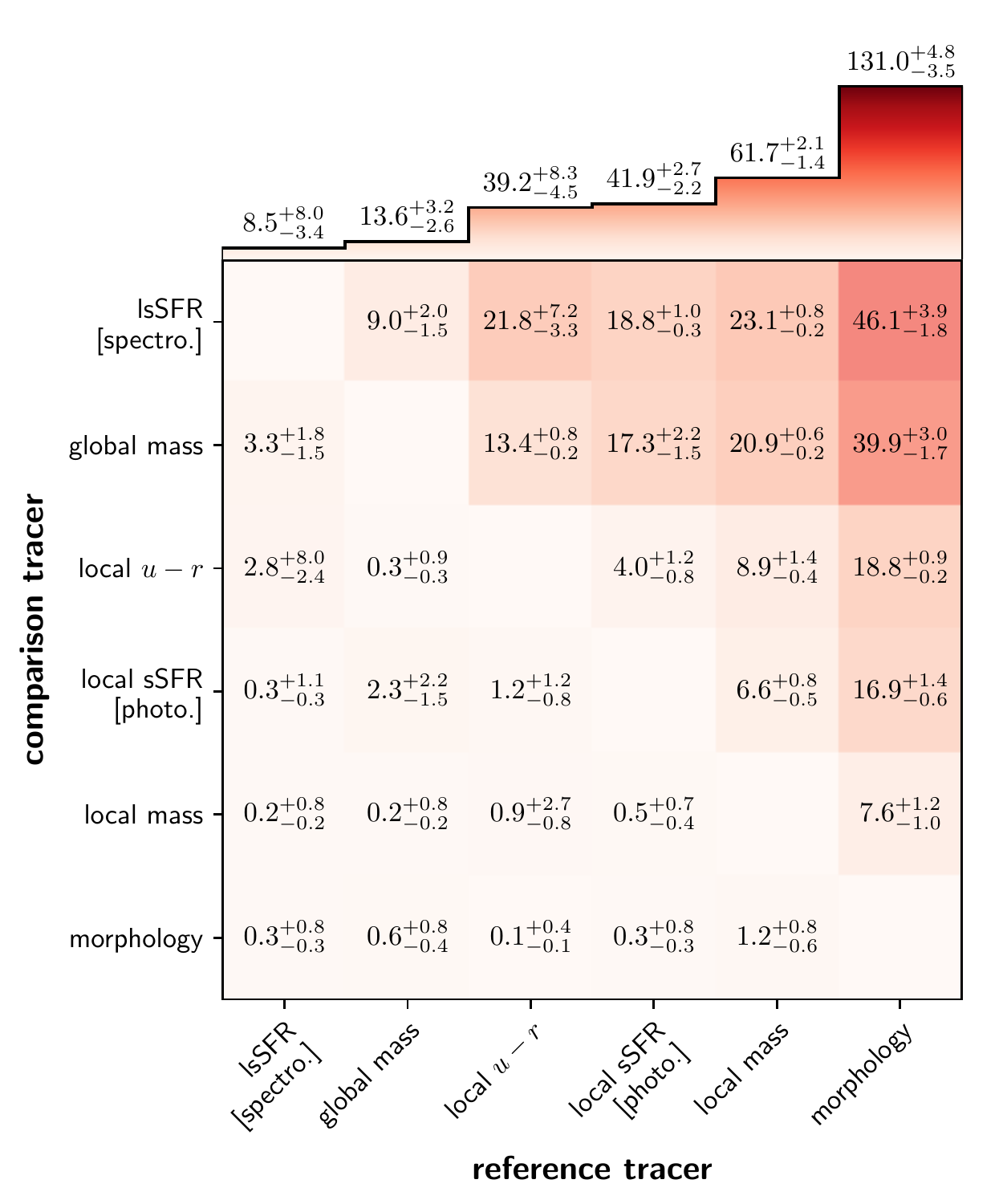}
	\caption{Matrix reporting the median $\chi^2$ for each reference tracer (\textit{columns}) and each comparison tracer (\textit{rows}) together with the 16\% and 84\% variations. The red color refers to the individual $\chi^2$ values: the darkest, the worst. On top of the matrix are plotted, for each reference tracer choice, the median with the 16\% and 84\% variations when summing the tracer walkers following Eq.~\ref{eq:chi2}.}
	\label{fig:chi2_matrix}
\end{figure}

Once we have determined the scatter $\sigma(c)$ expected given the amount of contamination, $c$, we can measure the $\chi^{2}$ associated to the ability of each reference tracer to explain the data, such that:
\begin{equation}
    \chi^2_\textrm{ref} = \sum_t \left(\frac{\gamma^*_t - \gamma(c_t)}{\sigma(c_t)}\right)^2
    \label{eq:chi2}
\end{equation}
where $t$ refers to the comparison tracers, $\gamma_t^*$ is the fitted step value fixing $\alpha$ and $\beta$ to these of the reference (see Section~\ref{sec:alphabeta_of_lssfr}) and $\gamma(c_t)$ is the expected step at contamination $c_t$ following Eq.~\ref{eq:stepbias}. Finally, since the $c_t$ measurements are noisy, we compute the $\chi^2_\textrm{ref}$ for each $c_t$ chain walkers. We report in Fig.~\ref{fig:chi2_matrix} (top panel) the median $\chi^2_\textrm{ref}$ for each tracer used as reference together with the 16\% and 84\% variations. This figure also displays the individual $\chi^2$ contributions (main panel).

The $\chi^2$ results confirm the qualitative observations. The spectroscopic lsSFR is the optimal reference tracer, followed by the global mass, the photometric lsSFR and the local $u-r$ colors. Local mass and morphology are the least suited. Since the spectroscopic lsSFR is the best reference tracer choice, it means that it is the most accurate to discriminate the underlying two populations, which further strengthens the claim that there seems to be the prompt vs. delay age dichotomy.

Quantitatively, with $\chi^2 = 8.5$ for 5 degrees of freedom, the scatter along the contamination line for spectroscopic lsSFR as the reference is consistent with random scatter of $1.1 \sigma$. The scatter for global mass as the reference has $\chi^2=13.6$, corresponding to a scatter of $2.1 \sigma$. All other tracers are excluded from being accurate reference tracer at more than $5 \sigma$.

\subsection{Scatter in the two-populations model}
\label{sec:scatter}

The two-populations model also has consequences for the observed scatter of the studied quantity $q$; see the introduction of the two-populations model in Section~\ref{sec:concept}. If both populations differ by $\gamma_0$ on average in $q$, and keeping the assumption of 50\% targets belonging to the $a$ population, then one can show that marginalizing the populations results in an additional scatter in the dispersion of $q$ by $0.5\times\gamma_0$.

In the context of SNe~Ia cosmology the studied quantify $q$ is the standardized magnitude and interestingly the intrinsic scatter, corresponding to the part of the standardized magnitude dispersion along the Hubble Diagram that cannot be explained by known sources of errors, typically is of $0.10\,\mathrm{mag}$ \citep{betoule2014,scolnic2018}. 

\begin{figure}
    \centering
    \includegraphics[width=\linewidth]{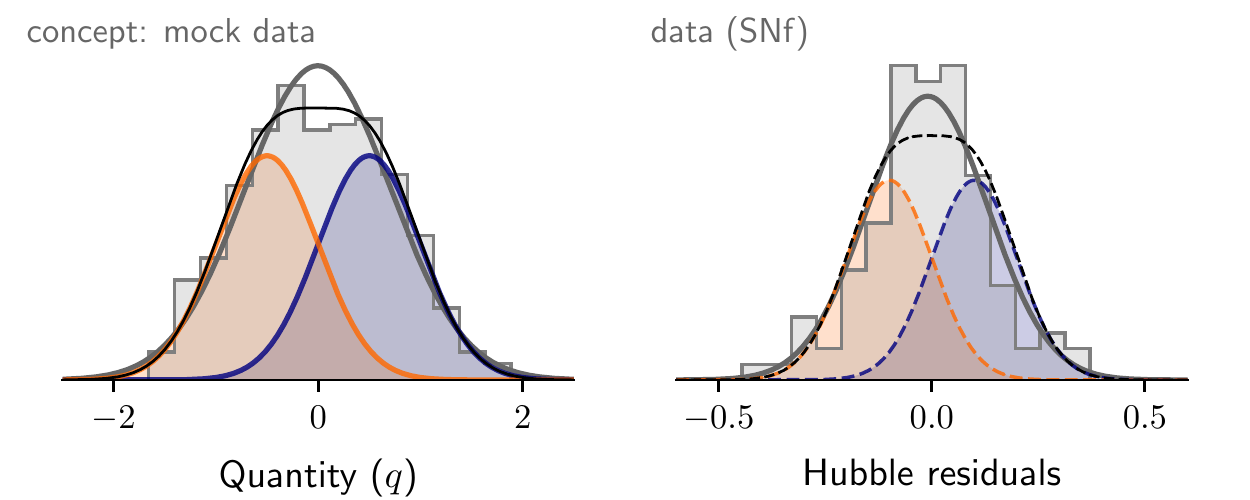}
	\caption{\textit{Left panel}: distribution of quantity $q$ for the same mock data as in top panel in Fig.~\ref{fig:contconcept}, represented by the grey histogram. The full thick grey line shows Gaussian parameters associated to this distribution. The blue and orange curves represent the estimation of the $q$ distribution from $a$ and $b$ populations respectively, while the black curve is the sum of both.
	\textit{Right panel}: distribution of the Hubble residuals of the SNfactory dataset, represented by the grey histogram. The full thick grey line shows the Gaussian that parameterizes this distribution. The dashed orange and dashed blue curves show the expected distribution of the two underlying populations that would explain the full intrinsic distributions (see Section~\ref{sec:scatter}). The dashed grey line is the sum of both and should be compared to the histogram.}
	\label{fig:hr_scatter}
\end{figure}

Hence, assuming the two-populations model, if the $0.10$~mag intrinsic SNe~Ia scatter were fully caused by the existence of two underlying populations, the average standardized magnitude difference between these populations would be $0.2$~mag. This is illustrated in the left panel of Fig.~\ref{fig:hr_scatter}: in this mock example, the observed full distribution of $q$ seems like a flattened distribution with a larger scatter than the underlying individual Gaussian distributions. 

If we apply this concept to the SNfactory dataset that also has an intrinsic dispersion of $\sim0.10$~mag \citep{rigault2020}, we can then guess the two underlying population distributions that would cause this effect; this is illustrated in the right panel of Fig.~\ref{fig:hr_scatter}. In the case of the SNfactory standardisation SNe~Ia distribution, the central part (mag $\sim0$) does not seem to qualitatively follow the expected distribution. This suggest that entire intrinsic distribution might not be fully explained by existence of a magnitude bias of $\sim0.2$~mag between two underlying SN~Ia populations.

Interestingly, this conceptual analysis provides a key information on the upper limit of the astrophysical bias affecting SNe~Ia standardized magnitudes in the context of the two populations model: it cannot be larger than twice the intrinsic dispersion, and, consequently, it is smaller than $\sim0.2$~mag. 

In addition, if the magnitude step related to the spectroscopic lsSFR is $\sim0.16$~mag, as claimed by \cite{rigault2020}, then the contamination of this tracer is lower than $25\%$.

\section{Conclusion}
\label{sec:conclusion}
We use a sample of 110 SNe~Ia from the Nearby Supernovae factory dataset to study the apparent inconsistencies in the literature between the different observed environmental dependencies of the standardized SNe~Ia magnitudes. In the last ten years, the SNe~Ia luminosity has been shown to significantly depend on host properties, ranging from barely significant variations when split by host galaxy morphology \citep[e.g.,][]{pruzhinskaya2020}, to a very significant $15\%$ luminosity difference when the SNe are split with respect to the spectroscopic specific star formation rate of their local environment \citep{rigault2020}; leaving the 8\% luminosity difference measured using the commonly-used global host galaxy mass step in between \citep[e.g.,][]{sullivan2010,roman2018}.

To study these variations we first analyse the mathematical implications of assuming a step function, i.e. of comparing the SN~Ia magnitude means when splitting the data into two bins. We show that doing so implicitly assumes two things: (1) that there exists two underlying populations that differ in standardized magnitudes and (2) that the environmental tracer used is somewhat able to distinguish them. Exploration of the implications of this implicit “two-populations“ model enables us to demonstrate that the expected step observed by a tracer depends on its ability to accurately discriminate the two underlying populations. In detail, if we call $c$ the fraction of targets misclassified by the environmental tracer, and if we call $\gamma_0$ the true difference in mean magnitudes between the two populations, then the expected measured magnitude means offset $\gamma$ is given by $\gamma=\gamma_0 \times \left(1-c\right)$. The higher the contamination, hence the lower the tracer accuracy, the lower the expected measured step. In addition, the intrinsic magnitude dispersion caused by marginalizing the populations is half their magnitude offset. Since SNe~Ia intrinsic dispersion typically is $\sim0.10\,\mathrm{mag}$, the upper limit magnitude offset between the two SNe~Ia populations would be $\sim0.20\,\mathrm{mag}$, if the entire SNe~Ia intrinsic scatter was caused by the existence of these two populations.

In light of that prediction, we derive the main literature environmental tracers for each of the 110 SNe~Ia, namely: spectroscopic and photometric measurements of the local specific star formation rate, the global and local stellar masses, the host galaxy morphology and the local color. In this first analysis, we assume that one of these tracers is set as a reference. This provides a lower limit on the expected true amplitude of $\gamma_0$.
We draw from this analysis the following conclusions:

\paragraph{Tracer contamination model:}
Our model of a ``two SN~Ia populations model observed with tracers of various accuracy'' explains well the observed variations. In Fig.~\ref{fig:envdiaglssfr} we show the expected vs. measured magnitude step as a function of the derived tracer contaminations and we find good agreement, supporting a two-populations model. When applied to the steps reported in the literature, our model is able to explain the observed variations.

\paragraph{Spectroscopic lsSFR as reference tracer:}
When compared to the other tracers, using the spectroscopic local specific star formation rate as a reference tracer can explain all other observations with a scatter at $1.1\sigma$. All other measurements are excluded as suitable reference tracers, with the possible exception of global mass, which shows a $2.1\sigma$ scatter, as we can see in Fig.~\ref{fig:tracer_envdiags} and quantified in Fig.~\ref{fig:chi2_matrix}.

\paragraph{The ``prompt vs. delayed'' model:}
The spectroscopic lsSFR measures the fraction of young stars in the SNe~Ia vicinity. As all observations are explained by using it as the reference tracer, the ``prompt vs. delayed'' progenitor age model seems to best represent the behavior of the underlying populations. \cite{nicolas2021} further show that this model also explains the observed redshift-drift of the SN~Ia stretch distribution.

\paragraph{Origin of the mass-step:}
It seems that two populations related to progenitor age, combined with tracer accuracy, can explain all previous measurements of the mass-step. This conclusion is in agreement with former analyses \cite[e.g.][]{rigault2013,rigault2020,roman2018}.

\paragraph{Standardisation coefficients $\alpha$, $\beta$:}
Because the standardising parameters such as stretch and color are correlated with the underlying two populations, hence with their tracers, the use of an inaccurate tracer, such as morphology, biases the derivation of $\alpha$ and $\beta$, as would a non-simultaneous estimation of $\alpha$, $\beta$ and $\gamma$.

\paragraph{The amplitude of $\gamma$:}
Under the assumption of the two-populations model, the amplitude of the astrophysical bias affecting the SNe~Ia luminosity (i.e. the intercept of the $\gamma(c)$ plot) is close to the age-step reported in \citet[][]{rigault2020} ($0.162\pm0.029$~mag), since the spectroscopic lsSFR is a good reference tracer. When we fit the intercept jointly on all literature data points using the derived contaminations from SNfactory sample, we find $0.121\pm0.010$ (see Fig.~\ref{fig:envdiaglssfr}).

\vspace{15pt}

In light of the described two-populations model and the importance of tracer accuracy when assessing the amplitude of the astrophysical bias in SNe~Ia cosmology, we highlight the importance of careful analyses of astrophysical biases when deriving cosmological parameters. This is true even when comparing two SN~Ia samples at similar redshift ranges, if their selection function would favor a given underlying population for any reason. 

To avoid biases, one might want to probe as accurately as possible the underlying populations and be careful when assessing them using only moderately good tracers such as global ones. In practice we raise warning for current cosmological analyses that use the mass-step as the third standardisation parameter to account for astrophysical dependencies in the SN~Ia magnitude. The host stellar mass is not the underlying parameter affecting the SN~Ia progenitor explosion mechanism or the way we see it, rather, it is a tracer correlated to the true underlying physics. As astrophysical properties evolve significantly with cosmic time, it is critical to understand the relationship between SN~Ia luminosity and the environment when doing SNe~Ia cosmology.

\begin{acknowledgements}
We thank the anonymous referee for the constructive comments which helped to improve the conclusions of the paper.
This project has received funding from the European Research Council (ERC) under the European Union's Horizon 2020 research and innovation program (grant agreement n$^\circ$759194 - USNAC).
\end{acknowledgements}

\bibliographystyle{aa} 

\onecolumn
\appendix
\section{Mathematical derivation of the modelisation for two populations of SNe~Ia}
\label{app:mathematical_derivation}
\subsection{Two population model}
The two population model estimates the probability of measuring certain fractions of false positives and false negatives, given known fractions of intrinsic false positives and false negatives.
Closely following the notations presented in Section~\ref{sec:concept}, we introduce the variables:
\begin{itemize}
    \item $k_i = \{a; b\}$ is a discrete indicator describing the true type of target $i$ given the two SN~Ia populations;
    \item $c_a$ and $c_b$ are, respectively, fractions of intrinsic false $a$ and false $b$ targets;
    \item $\hat{t}_i$ is the true tracer value for $i$ target; 
    \item $t_i$ is the measurement of tracer for $i$ target; 
    \item $\delta t_i$ is the measurement uncertainty of tracer for $i$ target;
    \item $t_\textrm{cut}$ is the cut-off value, discriminating the two categories, for tracer $t$.
\end{itemize}

Within this context, the probability to measure $t_i$ while the target $i$ truly belongs to $b$ population, given $c_b$, is expressed by:
\begin{align}
    \label{eq:alexfull}
    \mathcal{P}(k_i, t_i \mid c_a, c_b) &= \int_{-\infty}^{+\infty} \textrm{d}\hat{t}_i \int \textrm{d}\theta \; \mathcal{P}(k_i, t_i, \hat{t}_i, \theta \mid c_a, c_b) \nonumber\\
    &= \int_{-\infty}^{+\infty} \textrm{d}\hat{t}_i \int \textrm{d}\theta \; \mathcal{P}(t_i \mid k_i, \hat{t}_i, \theta, c_a, c_b) \, \mathcal{P}(k_i, \hat{t}_i, \theta \mid c_a, c_b) \nonumber\\
    &= \int_{-\infty}^{+\infty} \textrm{d}\hat{t}_i \int \textrm{d}\theta \; \mathcal{P}(t_i \mid k_i, \hat{t}_i, \theta, c_a, c_b) \, \mathcal{P}(\hat{t}_i \mid k_i, \theta, c_a, c_b) \, \mathcal{P}(k_i, \theta \mid c_a, c_b) \nonumber\\
    &= \int_{-\infty}^{+\infty} \textrm{d}\hat{t}_i \int \textrm{d}\theta \; \mathcal{P}(t_i \mid k_i, \hat{t}_i, \theta, c_a, c_b) \, \mathcal{P}(\hat{t}_i \mid k_i, \theta, c_a, c_b) \, \mathcal{P}(k_i \mid \theta, c_a, c_b) \, \mathcal{P}(\theta \mid c_a, c_b)
\end{align}
where $\theta$ captures all of the other model parameters that may exist, i.e., the nuisance parameters. The first of the four terms in the last integral is related to measurement uncertainties, second is the tracer probability, third is the type probability, the forth is the probability of drawing a target of class $k$ and the rest is how nuisance parameters are related to $c_a$ and $c_b$.

In this paper, we make the following assumptions:
\begin{enumerate}
    \item Knowledge of $\hat{t}_i$ is all that is needed to get $t_i$, so $\mathcal{P}(t_i \mid k_i, \hat{t}_i, \theta, c_a, c_b = \mathcal{P}(t_i \mid \hat{t}_i) = \mathcal{N}(t_i; \hat{t}_i, \delta t_i)$.
    \item The unknown underlying distribution of $\hat{t}_i$ only depends on the SN~Ia population type $k_i$ and the fraction of false $a$ or false $b$ targets, so $\mathcal{P}(\hat{t}_i \mid k_i=a, \theta, c_a) = \mathcal{P}(\hat{t}_i \mid k_i=a, c_a)$ and $\mathcal{P}(\hat{t}_i \mid k_i=b, \theta, c_b) = \mathcal{P}(\hat{t}_i \mid k_i=b, c_b)$.
    \item For we are only interested to know if a target is measured above or below a given cut, we use simple normalized top-hats ($\mathcal{U}$) to build the $t$ probability distribution functions, such that $\mathcal{P}(\hat{t}_i \mid k_i=b, c_b) = (1-c_b)\,\mathcal{U}(t_\textrm{min},t_\textrm{cut}) + c_b\,\mathcal{U}(t_\textrm{cut},t_\textrm{max})$ and $\mathcal{P}(\hat{t}_i \mid k_i=a, c_a) = c_a\,\mathcal{U}(t_\textrm{min},t_\textrm{cut}) + (1-c_a)\,\mathcal{U}(t_\textrm{cut},t_\textrm{max})$, where $(t_\textrm{min},t_\textrm{max})$ corresponds to boundaries for the parameter $\hat{t}_i$, though their values do not affect the inference (we use $t_\textrm{cut} - t_\textrm{min} = t_\textrm{max} - t_\textrm{cut} \gg \delta t_i$). The $\mathcal{U}$ are normalized such that $\mathcal{U}(x ; \mathrm{min}, \mathrm{max}) = (\mathrm{max}-\mathrm{min})^{-1}$ if x within min and max and 0 otherwise.
    \\
    We have tested the impact on this hypothesis on our results by simulating many mock samples with non-top hat functions;  namely Gaussian or Gaussian mixtures with various parameters. 
    When fitting these simulations with out baseline top-hat model, we accurately recover the input $c_a$ and $c_b$ values for a large range of $c$ values combinations.
    \item In first approximation, the fraction of true $a$ targets is constant. Notably, this requires that this fraction does not depend on redshift. This assumption, while most likely over simplistic for the general case, seems reasonable as we are studying data within a small redshift range ($0.03<z<0.08$). This results in $\mathcal{P}(k_i=a \mid \theta, c_a) = \mathcal{P}(k_i=a) = p_a$ and $\mathcal{P}(k_i=b) = (1-p_a)$.
    \item As a consequence of the given assumptions, there are no nuisance parameters ($\theta$) in the model. 
\end{enumerate}

This way, for the case of $k_i = b$, equation \ref{eq:alexfull} simplifies to:
\begin{align}
    \mathcal{P}(k_i=b, t_i \mid c_b) &= \int_{-\infty}^{+\infty} \textrm{d}\hat{t}_i \int \textrm{d}\theta \; \mathcal{P}(t_i \mid \hat{t}_i) \, \mathcal{P}(\hat{t}_i \mid k_i, c_b) \, \mathcal{P}(k_i) \, \mathcal{P}(\theta) \nonumber\\
    &= \mathcal{P}(k_i=b) \int_{-\infty}^{+\infty} \textrm{d}\hat{t}_i \; \mathcal{P}(t_i \mid \hat{t}_i) \, \mathcal{P}(\hat{t}_i \mid k_i=b, c_b) \nonumber\\
    &= (1-p_a) \int_{-\infty}^{+\infty} \textrm{d}\hat{t}_i \; \mathcal{N}(t_i; \hat{t}_i, \delta t) \, \Big((1-c_b)\,\mathcal{U}(t_\textrm{min},t_\textrm{cut}) + c_b\,\mathcal{U}(t_\textrm{cut},t_\textrm{max})\Big) \nonumber\\
    &= (1-p_a) \left(\int_{-\infty}^{t_\textrm{cut}} \textrm{d}\hat{t}_i \; \mathcal{N}(\hat{t}_i; t_i, \delta t) \, (1-c_b)\,\mathcal{U}(t_\textrm{min},t_\textrm{cut}) + \int_{t_\textrm{cut}}^{+\infty} \textrm{d}\hat{t}_i \; \mathcal{N}(\hat{t}_i; t_i, \delta t) \, c_b\,\mathcal{U}(t_\textrm{cut},t_\textrm{max})\right) \nonumber\\
    &= \frac{1-p_a}{\Delta} \left((1-c_b) \, \int_{t_\textrm{min}}^{t_\textrm{cut}} \textrm{d}\hat{t}_i \; \mathcal{N}(\hat{t}_i; t_i, \delta t) + c_b \, \int_{t_\textrm{cut}}^{t_\textrm{max}} \textrm{d}\hat{t}_i \; \mathcal{N}(\hat{t}_i; t_i, \delta t)\right) \nonumber\\
    &= \frac{1-p_a}{\Delta}\,\Big((1-c_b) \times f_i + c_b \times (1 - f_i)\Big)
\end{align}
where $f_i = \int_{t_\textrm{min}}^{t_\textrm{cut}} \textrm{d}\hat{t}_i \; \mathcal{N}(\hat{t}_i; t_i, \delta t)$ is assimilated to the cumulative distribution function (see eq.~\ref{eq:fi} in the main text) ; $\Delta =(t_\textrm{max}-t_\textrm{cut})=(t_\textrm{cut}-t_\textrm{min})$ is a constant normalisation term.
\subsection{Two population model with a reference tracer}
Adding the reference tracer (and applying the same assumptions as above), we get:
\begin{align}
    \mathcal{P}(k_i=b, t_i, t_i^\textrm{\:ref} \mid c_b, c_b^\textrm{\:ref}) &= \int_{-\infty}^{+\infty} \textrm{d}\hat{t}_i \int_{-\infty}^{+\infty} \textrm{d}\hat{t}_i^\textrm{\:ref} \int \textrm{d}\theta \; \mathcal{P}(k_i, t_i, \hat{t}_i, t_i^\textrm{\:ref}, \hat{t}_i^\textrm{\:ref}, \theta \mid c_b, c_b^\textrm{\:ref}) \nonumber\\
    &= \int_{-\infty}^{+\infty} \textrm{d}\hat{t}_i \int_{-\infty}^{+\infty} \textrm{d}\hat{t}_i^\textrm{\:ref} \int \textrm{d}\theta \;
    \begin{aligned}[t]
        & \; \mathcal{P}(t_i \mid k_i, \hat{t}_i, t_i^\textrm{\:ref}, \hat{t}_i^\textrm{\:ref}, \theta, c_b, c_b^\textrm{\:ref}) \\ 
        \times & \; \mathcal{P}(t_i^\textrm{\:ref} \mid k_i, \hat{t}_i, \hat{t}_i^\textrm{\:ref}, \theta, c_b, c_b^\textrm{\:ref}) \\ 
        \times & \; \mathcal{P}(\hat{t}_i \mid k_i, \hat{t}_i^\textrm{\:ref}, \theta, c_b, c_b^\textrm{\:ref}) \\ 
        \times & \; \mathcal{P}(\hat{t}_i^\textrm{\:ref} \mid k_i, \theta, c_b, c_b^\textrm{\:ref}) \\ 
        \times & \; \mathcal{P}(k_i \mid \theta, c_b, c_b^\textrm{\:ref}) \\ 
        \times & \; \mathcal{P}(\theta \mid c_b, c_b^\textrm{\:ref})
    \end{aligned}
    \nonumber\\
    &= \int_{-\infty}^{+\infty} \textrm{d}\hat{t}_i \int_{-\infty}^{+\infty} \textrm{d}\hat{t}_i^\textrm{\:ref} \int \textrm{d}\theta \; \mathcal{P}(t_i \mid \hat{t}_i) \, \mathcal{P}(t_i^\textrm{\:ref} \mid \hat{t}_i^\textrm{\:ref}) \, \mathcal{P}(\hat{t}_i \mid k_i, c_b) \, \mathcal{P}(\hat{t}_i^\textrm{\:ref} \mid k_i, c_b^\textrm{\:ref}) \, \mathcal{P}(k_i) \, \mathcal{P}(\theta) \nonumber\\
    &= \mathcal{P}(k_i=b) \int_{-\infty}^{+\infty} \textrm{d}\hat{t}_i \int_{-\infty}^{+\infty} \textrm{d}\hat{t}_i^\textrm{\:ref} \; \mathcal{P}(t_i \mid \hat{t}_i) \, \mathcal{P}(t_i^\textrm{\:ref} \mid \hat{t}_i^\textrm{\:ref}) \, \mathcal{P}(\hat{t}_i \mid k_i=b, c_b) \, \mathcal{P}(\hat{t}_i^\textrm{\:ref} \mid k_i=b, c_b^\textrm{\:ref}) \nonumber\\
    &= (1 - p_a) 
    \begin{aligned}[t]
        & \times \int_{-\infty}^{+\infty} \textrm{d}\hat{t}_i \; \mathcal{N}(t_i; \hat{t}_i, \delta t) \, \Big((1-c_b)\,\mathcal{U}(t_\textrm{min},t_\textrm{cut}) + c_b\,\mathcal{U}(t_\textrm{cut},t_\textrm{max})\Big) \\
        & \times \int_{-\infty}^{+\infty} \textrm{d}\hat{t}_i^\textrm{\:ref} \; \mathcal{N}\Big(t_i^\textrm{\:ref}; \hat{t}_i^\textrm{\:ref}, \delta t^\textrm{\:ref}\Big) \, \bigg((1-c_b^\textrm{\:ref})\,\mathcal{U}\Big(t_\textrm{min}^\textrm{\:ref},t_\textrm{cut}^\textrm{\:ref}\Big) + c_b^\textrm{\:ref}\,\mathcal{U}\Big(t_\textrm{cut}^\textrm{\:ref},t_\textrm{max}^\textrm{\:ref}\Big)\bigg)
    \end{aligned}
    \nonumber\\
    &= \frac{1-p_a}{\Delta'}\, \Big((1-c_b) \times f_i + c_b \times (1 - f_i)\Big) \, \Big(\big(1-c_b^\textrm{\:ref}\big) \times f_i^\textrm{\:ref} + c_b^\textrm{\:ref} \times \big(1 - f_i^\textrm{\:ref}\big)\Big)
\end{align}

Similarly for $k_i = a$, we get:
\begin{equation}
    \mathcal{P}(k_i=a, t_i, t_i^\textrm{\:ref} \mid c_a, c_a^\textrm{\:ref}) = \frac{p_a}{\Delta'}  \, \Big(c_a \times f_i + (1-c_a) \times (1 - f_i)\Big) \, \Big( c_a^\textrm{\:ref} \times f_i^\textrm{\:ref} + \big(1-c_a^\textrm{\:ref}\big) \times \big(1 - f_i^\textrm{\:ref}\big)\Big)
\end{equation}

Finally, marginalizing over the population type, we find the general form of Eq.~\ref{eq:probatracercomplete}: 
\begin{align}
    \mathcal{P}(t_i, t_i^\textrm{\:ref} \mid c_a, c_b, c_a^\textrm{\:ref}, c_b^\textrm{\:ref}) &= \int \textrm{d}k_i \; \mathcal{P}(k_i, t_i, t_i^\textrm{\:ref} \mid c_a, c_b, c_a^\textrm{\:ref}, c_b^\textrm{\:ref}) \nonumber\\
    &= \mathcal{P}(k_i=a, t_i, t_i^\textrm{\:ref} \mid c_a, c_a^\textrm{\:ref}) + \mathcal{P}(k_i=b, t_i, t_i^\textrm{\:ref} \mid c_b, c_b^\textrm{\:ref}) \nonumber\\
    &\propto p_a \, \Big( c_a^\textrm{\:ref} \times f_i^\textrm{\:ref} + \big(1-c_a^\textrm{\:ref}\big) \times \big(1 - f_i^\textrm{\:ref}\big)\Big) \, \Big(c_a \times f_i + (1-c_a) \times (1 - f_i)\Big) \nonumber\\
    & \quad + (1-p_a) \, \Big(\big(1-c_b^\textrm{\:ref}\big) \times f_i^\textrm{\:ref} + c_b^\textrm{\:ref} \times \big(1 - f_i^\textrm{\:ref}\big)\Big) \, \Big((1-c_b) \times f_i + c_b \times (1 - f_i)\Big)
\end{align}

\subsection{Testing the model robustness over the assumptions}
We tested on simulations the two main assumptions made on this analysis ; namely, (1) that the fraction of class "a"  target is a constant equal to 50\%  and (2) that the pdfs of the tracer values can be approximated by combination of top-hat distributions.

\begin{enumerate}
    \item To test the impact of a fixed $p_a=50\%$ term, we have simulated many different models with varying $p_a$ values ranging from 1 to 99\%. Each time we built 5 different tracers with various $c_a$ and $c_b$ ranging from 5\% to 50\%, some symmetric ($c_a=c_b$), some not. When fitting these simulated samples assuming our baseline model (hence with $p_a=50\%$) we recovered the input $c_a$ and $c_b$ values with no bias as long as the input $p_a$ is within $25-75\%$ ; the bias become significant (at more than $3\sigma$) only on the extreme ($p_a<10\%$ or $p_a>90\%$).
    \item To test the assumption made on the tracer distribution in order to simplify the model, we have build many simulations while varying the assumed distributions. We used Gaussian or Gaussian mixture models with various parameter values to draw a perfect tracer prior to shuffling below (above) the tracer's cut a fraction $c_a$ ($c_b$) of target to simulate the tracers' contamination. Random noise is added next and we did so for many combinations of $c_a$ and $c_b$ values. When fitting these simulated samples with our baseline model, we each time recovered the input $c_a$ and $c_b$ values with no bias.
\end{enumerate}

We hence conclude that the assumptions made in the paper to simplify the likelihood have no consequences on our results.

\end{document}